\documentclass[journal]{IEEEtran}
\usepackage{blindtext}
\usepackage{graphicx}

\usepackage{bm}
\usepackage{amsfonts}

\usepackage{amssymb}

\usepackage{cite}

\ifCLASSINFOpdf
\else
\fi
\usepackage{amsmath}
\usepackage{algorithmic}

\usepackage{color}

\usepackage{stfloats}
\fnbelowfloat
\begin{document}
%
\title{Interference Exploitation for Radar and Cellular Coexistence: The Power-Efficient Approach}
%
%
%

\author{Fan Liu,~\IEEEmembership{Student~Member,~IEEE,}
        Christos Masouros,~\IEEEmembership{Senior~Member,~IEEE,}
        Ang Li,~\IEEEmembership{Student~Member,~IEEE,}
        Tharmalingam Ratnarajah,~\IEEEmembership{Senior~Member,~IEEE,}
        and~Jianming Zhou
\thanks{Manuscript received ***}
\thanks{Fan Liu and Jianming Zhou are with the School of Information and Electronics, Beijing Institute of Technology, Beijing, 100081, China (e-mail: liufan92@bit.edu.cn, zhoujm@bit.edu.cn). 
Fan Liu, Christos Masouros and Ang Li are with the Department of Electronic and Electrical Engineering, University College London, London, WC1E 7JE, UK (e-mail: ang.li.14@ucl.ac.uk, chris.masouros@ieee.org). 
Tharmalingam Ratnarajah is with the Institute for Digital Communications, School of Engineering, The University of Edinburgh, Edinburgh, EH9 3JL, UK (e-mail: t.ratnarajah@ed.ac.uk).
}
}

\maketitle

\begin{abstract}
We propose a novel approach to enable the coexistence between Multi-Input-Multi-Output (MIMO) radar and downlink multi-user Multi-Input-Single-Output (MU-MISO) communication system. By exploiting the constructive multi-user interference (MUI), the proposed approach trades-off useful MUI power for reducing the transmit power, to obtain a power efficient transmission. This paper focuses on two optimization problems: a) Transmit power minimization at the base station (BS) while guaranteeing the receive signal-to-interference-plus-noise ratio (SINR) level of downlink users and the interference-to-noise ratio (INR) level to radar; b) Minimization of the interference from BS to radar for a given requirement of downlink SINR and transmit power budget. To reduce the computational overhead of the proposed scheme in practice, an algorithm based on gradient projection is designed to solve the power minimization problem. In addition, we investigate the trade-off between the performance of radar and communication, and analytically derive the key metrics for MIMO radar in the presence of the interference from the BS. Finally, a robust power minimization problem is formulated to ensure the effectiveness of the proposed method in the case of imperfect Channel State Information (CSI). Numerical results show that the proposed method achieves a significant power saving compared to conventional approaches, while obtaining a favorable performance-complexity trade-off.
\end{abstract}


%
\IEEEpeerreviewmaketitle

\section{Introduction}
%
%
%
%
\IEEEPARstart {I}{n} response to the increasing demand for wireless communication devices and services, the Federal Communications Commission (FCC) has adopted a broadband plan to release an additional 500MHz spectrum that is currently occupied by military and governmental operations, such as air surveillance and weather radar systems\cite{federal2010connecting}. Since then, spectrum sharing between radar and communication has been regarded as an enabling solution. In \cite{7131098}, a radar information rate has been defined, such that the performance of radar and communication can be discussed using the same metric. Similar work has been done in \cite{6875553,7279172}, in which radar and communication are unified under the framework of information theory, and the channel capacity between radar and target has been defined by applying the rate distortion theory. Nevertheless, these works focus on the single-antenna systems rather than MIMO systems. At present, several methods considering the spectrum sharing between MIMO radar and communication have been proposed \cite{6503914,4058251,7347464,6735841,7485316,7089157,6831613,6636787,7127473,7289385,6956861,7485158, 6933960, 7485066,7470514, 7418294, 7178410}, since traditional radar will soon be replaced by MIMO radar in the near future due to the advantages of waveform diversity and higher detection capability\cite{4350230,li2009mimo}. In \cite{4058251}, the feasibility of combining MIMO radar and Orthogonal Frequency Division Multiplexing (OFDM) communication has been studied. More recently, a novel dual-functional waveform has been reported by \cite{7347464}, where communication bits are embedded in the radar waveform by controlling the sidelobe of the transmit beam patterns for radar. Other related schemes, including modulating information by shuffling the waveform across the radar transmit antennas and Phase Shift Keying (PSK) modulation via different weight vectors, have been proposed in \cite{7485316,7485066 }.
\\\indent More relevant to this work, transmit beamforming has been viewed as a promising solution to eliminating the mutual interference between radar and communication. First pioneered by\cite{6503914}, the idea of null space projection (NSP) beamforming has been widely discussed \cite{7089157,6831613,6636787,7127473,7289385,6956861}, where the radar waveforms are projected onto the null space of the interference channel matrix from radar transmitter to communication receiver. Optimization-based beamforming has been exploited to solve the problem in \cite{7485158}, where the SINR of radar has been optimized subject to power and capacity constraints of communication. Related work discusses the coexistence between MIMO-Matrix Completion (MIMO-MC) radar and MIMO communication system, where the radar beamforming matrix and communication covariance matrix are jointly optimized \cite{7470514}. In general, existing works on interference mitigation for radar-communication coexistence mainly consider the scenario between MIMO radar and point-to-point MIMO (P2P MIMO) communication, while few efforts have been taken for the case of radar and multi-user communication. Moreover, none of above works discusses the case of imperfect CSI.
\\\indent Motivated by the robust beamforming in the broader area of cognitive radio networks \cite{4787135,6373750}, the work \cite{liu2016robust} investigated the robust MIMO beamforming for the coexistence of radar and downlink MU communication, where the radar detection probability was maximized while guaranteeing the transmit power of BS and the receive SINR for each downlink user using Semidefinite Relaxation (SDR) techniques\cite{5447068,5447076}. In such optimizations, all the interference from other downlink users is regarded as harmful to the user of interest. Nevertheless, previous works proved that for a downlink MU-MIMO system using PSK modulations, the known interference can act constructively to benefit the symbol decision at downlink users. In \cite{4801492}, partial channel inversion was applied to the BS such that the constructive part of MUI was preserved while the destructive part was eliminated. Further research \cite{5159472,5605266} reported that by rotating the interference into the direction of the signal of interest, the MUI was always kept constructive. Moreover, recent works\cite{6619580,7103338} showed that by rotating the destructive interference into constructive region using optimization techniques, the receive SINR target for each user was actually relaxed compared to the conventional SDR-based beamformer, thus a significant power saving was obtained. This work has also been applied to cognitive radio transmission to design closed-form precoding solutions\cite{6193461,6094141}.
\\\indent In this paper, we develop a novel precoding optimization approach for the spectrum sharing between MIMO radar and downlink MU-MISO communication based on the concept of constructive interference (CI). By allowing the BS to utilize the known interference as a green signal power, the receive SINR at the users is increased. In fact, for a given SINR constraint using constructive interference, the feasible domain of the optimization problem is extended compared to the conventional SDR-based beamforming. We consider two optimization-based transmit beamforming designs, one is to minimize the transmit power at the BS while guaranteeing the receive SINR at the users and the interference level from BS to radar, the other is to minimize the total interference from BS to radar subject to the SINR constraint per user and transmit power budget. It is worth noting that both problems are convex and can be optimally solved by numerical tools. To efficiently apply the proposed schemes in practice, we design an efficient gradient projection algorithm for power minimization by analyzing the structure of the optimization. To investigate the effect of interference minimization beamforming on the performance of radar, we further derive the analytic form of detection probability and Cram\'er-Rao bound (CRB) for MIMO radar with the presence of the interference from the BS. By doing so, important trade-offs between the performance of radar and communication are given. Finally, we consider the uncertainty in the estimated channel information, and design a worst-case robust beamformer based on the principle of interference exploitation. For clarity, we list the contributions of this paper as follows:

\begin{itemize}
\item We design a power efficient optimization-based beamforming technique for the coexistence of MIMO radar and downlink MU-MISO communication based on exploiting the constructive interference power, where two optimization problems are formulated: a) Power minimization subject to SINR and INR constraints; b) Interference minimization subject to SINR and power constraints. The proposed approach outperforms the conventional SDR-based method.
\item We investigate the structure of the power minimization problem, and derive a computationally efficient algorithm to solve it.
\item We analytically derive the detection probability and the CRB for MIMO radar when the proposed beamforming scheme is used.
\item We derive the robust beamforming design of power minimization for the case of imperfect CSI.
\end{itemize}

The remainder of this paper is organized as follows. Section II introduces the system model and briefly recalls the conventional SDR-based beamforming problems. Section III describes the concept of CI and formulates the proposed optimization problems using the CI technique. In Section IV, a thorough analysis for the power minimization optimization is present and an efficient algorithm is derived. Section V derives the detection probability and the Cram\'er-Rao bound of MIMO radar for the proposed scenario. A worst-case approach for imperfect CSI is given for robust power minimization in Section VI, with norm-bounded CSI errors. Numerical results are provided and discussed in Section VII. Finally, Section VIII concludes the paper.
\\\indent{\emph{Notations}}: Matrices are denoted by bold uppercase letters (i.e., ${\bf{H}}$), bold lowercase letters are used for vectors (i.e., $\pmb\beta$), subscripts indicate the rows of a matrix unless otherwise specified (i.e., ${\bf{h}}_i$ is the \emph{i}-th row of ${\bf{H}}$), scalars are denoted by normal font (i.e., $R_m$), $\text{tr}\left(\cdot\right)$ stands for the trace of the argument, $(\cdot)^T$, $(\cdot)^*$ and $(\cdot)^H$ stand for transpose, complex conjugate and Hermitian transpose respectively, $\operatorname{Re}(\cdot)$ and $\operatorname{Im}(\cdot)$ denote the real and imaginary part of the argument.

\section{System Model and SDR-based Beamforming}
Consider a spectrum sharing scenario where a \emph K-user MU-MISO downlink system operates at the same frequency band with a MIMO radar. As can be seen in Fig. 1, the \emph N-antenna BS is transmitting signals to \emph K single-antenna users while the MIMO radar with $M_t$ transmit antennas and $M_r$ receive antennas is detecting a point-like target in the far-field. Inevitably, these two systems will cause interference to each other. The received signal at the \emph i-th downlink user is given as
\begin{figure}
    \centering
    \includegraphics[width=3.0in]{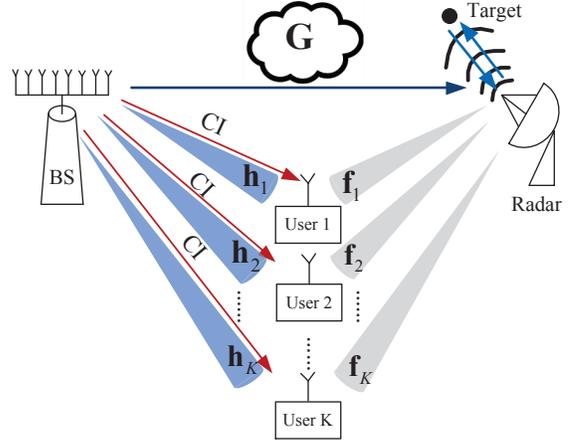}
    \caption{Spectrum sharing scenario.}
    \label{fig:1}
\end{figure}
\begin{equation}
    {{y}^C_i[l]} = {\bf{h}}_i^T\sum\limits_{k = 1}^K {{{\bf{t}}_k}{d}_k[l]}  + \sqrt{P_R}{\bf{f}}_i^T{{\bf s}_l} + {n}_i[l], i = 1,2,...,K,
\end{equation}
where $ {\bf{h}}_{i} \in {\mathbb{C}^{\emph{N} \times 1}} $ denotes the communication channel vector, $ {\bf{f}}_{i} \in {\mathbb{C}^{{{\emph{M}}_{t}} \times 1}} $ denotes the interference channel vector from radar to the user, $ {\bf{t}}_{i} \in {\mathbb{C}^{\emph{N} \times 1}} $ denotes the precoding vector, $ {d}_i[l] $ and $ {n}_i[l] \sim {\mathcal{C}}{\mathcal{N}}\left( {0,{\sigma _C^2}} \right)$ stands for the communication symbol and the received noise for the \emph{i}-th user. $ l = 1,2,...,L $ is the symbol index, $ L $ is the length of the communication frame, and $P_R$ is the power of radar signal. Without loss of generality, we assume that the communication symbol is drawn from a normalized PSK constellation, while we note that the proposed concept of interference exploitation has been shown to offer benefits for other modulation formats \cite{7417066,7831497}. Hence, the PSK symbol can be denoted as ${d_k}[l] = e ^ {j{\phi _k}[l]} $. It is assumed that $ {\bf{H}}=\left[{\bf{h}}_1, {\bf{h}}_2,...,{\bf{h}}_K \right] $ and $ {\bf{F}}=\left[{\bf{f}}_1, {\bf{f}}_2,...,{\bf{f}}_K \right] $ are flat Rayleigh fading and statistically independent with each other, and can be estimated by the BS through the pilot symbols.
\\\indent The second term at the right hand of (1) denotes the interference from radar to the user, where $ {\bf{S}} = \left[ {\bf{s}}_1,{\bf{s}}_2,...,{\bf{s}}_{L_R} \right] \in {\mathbb{C}^{M_t \times L_R}} $ is the radar transmit waveforms. According to the standard assumption in MIMO radar literature \cite{4350230,1703855}, ${\bf{S}}$ is set to be orthogonal, i.e.,  $\mathbb{E}\left[ {\bf s}_l{{\bf s}^H_l} \right] = \frac{1}{L_R}\sum\limits_{l = 1}^{L_R} {{\bf s}_l{{\bf s}^H_l}}  = {\bf{I}}$, where $\mathbb{E}$ denotes the ensemble average. For notational convenience, it is assumed that the symbol duration of the radar waveform is the same as the communication signal. It should be highlighted that in order to preserve the orthogonality of $\bf S$, radar may utilize codeword that is longer than a typical communication frame. Without loss of generality, we assume $L_R=L$ for the ease of our derivation.
\\\indent Based on the above, the receive SINR is given by
\begin{equation}
    {\gamma _i} = \frac{{{{\left| {{\bf{h}}_i^T{{\bf{t}}_i}} \right|}^2}}}{{\sum\limits_{k = 1,k \ne i}^K {{{\left| {{\bf{h}}_i^T{{\bf{t}}_k}} \right|}^2} + {P_R}{{\left\| {{{\bf{f}}_i}} \right\|}^2} + \sigma _C^2} }}, \forall i.
\end{equation}
And the average transmit power of the BS is 
\begin{equation}
    {P_C} = \sum\limits_{k = 1}^K {{{\left\| {{{\bf{t}}_k}} \right\|}^2}}.
\end{equation}
\\\indent With the presence of a point-like target located at direction $\theta$, the echo wave that received by radar at the \emph l-th time slot is
\begin{equation}
    {{\bf{y}}^R_l}=\alpha \sqrt{P_R} {\bf{A}}\left( \theta  \right){{\bf s}_l} + {{\bf{G}}^T}\sum\limits_{k = 1}^K {{{\bf{t}}_k}{d_k}\left[ l \right]} + {\mathbf{z}}_l,
\end{equation}
where  $ {\bf G }=\left[{\bf{g}}_1,{\bf{g}}_2,...,{\bf{g}}_{M_r} \right] \in {\mathbb{C}}^{N\times M_r}$ is the interference channel matrix between BS and radar RX, and is also assumed to be flat Rayleigh fading and statistically independent with other two channels, and is estimated at the BS, $\alpha \in \mathbb{C}$ is the complex path loss of the path between radar and target,  $ {\mathbf{z}}_l=\left[z_1\left[l\right], z_2\left[l\right],..., z_{M_r}\left[l\right]\right]^T \in {\mathbb{C}}^{M_r\times 1} $ is the received noise at the \emph{l}-th time slot with $z_m[l]\sim {\mathcal{C}}{\mathcal{N}}\left( {0,{\sigma _R^2}} \right), \forall m$, $ {\bf{A}}\left( \theta  \right) = {{\bf{a}}_R}\left( \theta  \right){\bf{a}}_T^T\left( \theta  \right) $, in which $ {{\bf{a}}_T}\left( \theta  \right) \in {\mathbb{C}^{{M_t} \times 1}} $ and $ {{\bf{a}}_R}\left( \theta  \right) \in {\mathbb{C}^{{M_r} \times 1}}  $ are transmit and receive steering vectors of the radar antenna array. The model in (4) is assumed to be obtained in a single range-Doppler bin of the radar detector and thus ignores the range and Doppler parameters. In this paper, we apply the basic assumptions in \cite{1703855} on the radar model, which is 
\begin{equation}
\begin{gathered}
  {M_r} = {M_t} = M, 
  \;\;{{\mathbf{a}}_R}\left( \theta  \right) = {{\mathbf{a}}_T}\left( \theta  \right) = {\mathbf{a}}\left( \theta  \right), \hfill \\
  {{\mathbf{A}}_{im}}\left( \theta  \right) = {{\mathbf{a}}_i}\left( \theta  \right){{\mathbf{a}}_m}\left( \theta  \right) = e^{ { - j\omega {\tau _{im}}\left( \theta  \right)} } \hfill \\
   \quad = e^ {\left( { - j\frac{{2\pi }}{\lambda }{{\left[ {\sin \left( \theta  \right);\cos \left( \theta  \right)} \right]}^T}\left( {{{\mathbf{x}}_i} + {{\mathbf{x}}_m}} \right)} \right)}, \hfill \\ 
\end{gathered} 
\end{equation}
where $ \omega $ and $ \lambda $ denote the frequency and the wavelength of the carrier, $ {{\bf A}_{im}}\left( \theta  \right) $ is the $ i $-th element at the $ m $-th column of the matrix $\bf A$, which is the total phase delay of the signal that transmitted by the \emph{i}-th element and received by the \emph{m}-th element of the antenna array, and $ {{\bf{x}}_i} = \left[ {x_i^1;x_i^2} \right] $ is the location of the \emph{i}-th element of the antenna array. In the above radar signal model, it is assumed that the communication interference is the only interference received by radar. Following the closely related literature, the interference caused by clutter and false targets is not considered\cite{7089157}. For convenience, we ignore the time index $ l $ in the rest of the paper unless otherwise specified. The interference from the BS on the \emph{m}-th antenna of radar is given by
\begin{equation}
   {u_m} = {\mathbf{g}}_m^T\sum\limits_{k = 1}^K {{{\mathbf{t}}_k}d_k}.
\end{equation}
We define the INR at the \emph{m}-th receive antenna of radar as
\begin{equation}
   {r_m} = \frac{{{{\left| {{u_m}} \right|}^2}}}{{\sigma _R^2}}. 
\end{equation}
\\\indent From a conventional perspective, all interference should be treated as harmful when optimizing the performance of the two systems. The power minimization problem of the BS subject to INR and SINR thresholds is formulated as 
\begin{equation}
\begin{array}{*{20}{l}}
  {{{\cal{P}}_0}:\mathop {\min }\limits_{{{\mathbf{t}}_k}} \;\;\sum\limits_{k = 1}^K {{{\left\| {{{\mathbf{t}}_k}} \right\|}^2}} {\text{  }}} \\ 
\displaystyle  {s.t.{\kern 1pt} {\kern 1pt} {\kern 1pt} {\kern 1pt} \frac{{{{\left| {{\mathbf{h}}_i^T{{\mathbf{t}}_i}} \right|}^2}}}{{\sum\limits_{k = 1,k \ne i}^K {{{\left| {{\mathbf{h}}_i^T{{\mathbf{t}}_k}} \right|}^2} + {P_R}{{\left\| {{{\mathbf{f}}_i}} \right\|}^2} + \sigma _C^2} }} \ge {\Gamma _i},\forall i,} \\ 
\displaystyle   {{\kern 1pt} {\kern 1pt} {\kern 1pt} {\kern 1pt} {\kern 1pt} {\kern 1pt} {\kern 1pt} {\kern 1pt} {\kern 1pt} {\kern 1pt} {\kern 1pt} {\kern 1pt} {\kern 1pt} {\kern 1pt} {\kern 1pt} \frac{{{{\left| {{\mathbf{g}}_m^T\sum\limits_{k = 1}^K {{{\mathbf{t}}_k}} {d_k} } \right|}^2}}}{{\sigma _R^2}} \le {R_m},\forall m,} \\  
\end{array}
\end{equation}
where $ \Gamma_i $ is the required SINR of the \emph{i}-th communication user, $ R_m $ is the maximum tolerable INR level of the \emph{m}-th receive element of radar. Similarly, we can formulate the optimization problem that minimizes the interference to radar while guaranteeing the BS power budget and the required SINR level at each user, which is given as
\begin{equation}
\begin{gathered}
  {{\cal{P}}_1}: \mathop {\min }\limits_{{{\mathbf{t}}_k}} {\kern 1pt} {\kern 1pt} \sum\limits_{m = 1}^M {{\kern 1pt} {\kern 1pt} {{\left| {{\mathbf{g}}_m^T\sum\limits_{k = 1}^K {{{\mathbf{t}}_k}} {d_k}} \right|}^2}}  \hfill \\
  s.t.{\kern 1pt} {\kern 1pt} {\kern 1pt} {\kern 1pt} {\kern 1pt} {\kern 1pt} \sum\limits_{k = 1}^K {{{\left\| {{{\mathbf{t}}_k}} \right\|}^2}}  \le P, \hfill \\
  \;\;\;\;\;\frac{{{{\left| {{\mathbf{h}}_i^T{{\mathbf{t}}_i}} \right|}^2}}}{{\sum\limits_{k = 1,k \ne i}^K {{{\left| {{\mathbf{h}}_i^T{{\mathbf{t}}_k}} \right|}^2} + {P_R}{{\left\| {{{\mathbf{f}}_i}} \right\|}^2} + \sigma _C^2} }} \ge {\Gamma _i},\forall i, \hfill \\ 
\end{gathered} 
\end{equation}
where $P$ is the budget of the BS transmit power. Problem ${{\cal{P}}_0}$ and ${{\cal{P}}_1}$ can be transformed into Semidefinite Program (SDP)\cite{boyd2004convex} with Semidefinite Relaxation techniques, and thus can be solved by numerical tools. We refer readers to \cite{liu2016robust,5447068,5447076} for more details on this topic. As shown in Fig. 1 by red arrows, it is worth noting the above problems ignore the fact that for each user, interference from other users can contribute to the received signal power constructively. In this paper, we aim to show that the solution of these problems is suboptimal from an instantaneous point of view and design a symbol-based beamforming method in accordance to the concept of constructive interference.

\section{Beamforming with Constructive Interference}
\begin{figure}
    \centering
    \includegraphics[width=2.8in]{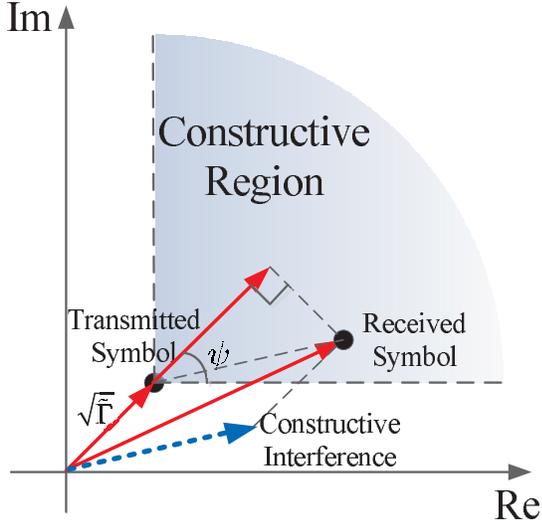}
    \caption{The principle of constructive interference.}
    \label{fig:2}
\end{figure}
As per the model of\cite{7103338}, the instantaneous interference can be divided into two categories, constructive interference and destructive interference. Generally, the constructive interference is defined as the interference that moves the received symbol away from the decision thresholds. The purpose of the CI-based beamforming is to rotate the known interference from other users such that the resultant received symbol falls into the constructive region. This is shown in Fig. 2, where we denote the constructive area of the QPSK symbol by the gray shade. It has been proven in\cite{7103338} that the optimization will become more relaxed than conventional interference cancellation optimizations due to the expansion of the optimization region. Hence, the performance of the beamformer is improved. Here we consider the instantaneous transmit power, which is given as
\begin{equation}
{P_T}[l] = {\left\| {\sum\limits_{k = 1}^K {{{\mathbf{t}}_k}e^{ {j\left({\phi _k}[l]-{\phi_1}[l]\right)} }} } \right\|^2},
\end{equation} 
where $d_1[l]=e^{j\phi_1[l]}$ is used as the phase reference. Based on \cite{7103338}, we rewrite the SINR constraints of ${{\cal{P}}_0}$ and ${{\cal{P}}_1}$ in a CI sense, and reformulate the power minimization problem ${{\cal{P}}_0}$ as (11) on the top of the next page as ${{\cal{P}}_2}$, where $\psi=\frac{\pi}{M_p}$, and $M_p$ is the PSK modulation order. Readers are referred to \cite{7103338} for a detailed derivation of the CI constraints and classification. It should be highlighted that, while here we focus on PSK constellations, the optimizations ${\cal{P}}_2$ onwards can be readily adapted to other constellation formats such as Quadrature Amplitude Modulation (QAM)\cite{7417066,7831497}. Note that ${{\cal{P}}_2}$ is convex in contrast to the non-convex ${{\cal{P}}_0}$ and ${{\cal{P}}_1}$. To be more specific, problem ${{\cal{P}}_2}$ is a second-order cone program (SOCP) and can be solved optimally by numerical tools.  
\\\indent In both ${{\cal{P}}_0}$ and ${{\cal{P}}_2}$, by letting $R_m = 0$, it follows $ {{\mathbf{g}}_m^T\sum\limits_{k = 1}^K {{{\mathbf{t}}_k}} {d_k} } =0 $, which requires the transmitting signal to fall into the null space of the interference matrix $\bf G$ and causes zero interference to radar. This yields the solution with which the radar can achieve the best performance. However, the strict equality will result in a large transmit power at BS. On the other hand, if we let $R_m \to \infty$, the INR constraints will be ineffective, which is equivalent to the typical downlink power minimization in the absence of radar. This trade-off between radar and communication performance will be further evaluated by numerical simulations below.

\begin{figure*}[!t]
\normalsize
\newcounter{MYtempeqncnt}
\setcounter{MYtempeqncnt}{\value{equation}}
\setcounter{equation}{10}
\begin{equation}
\begin{array}{*{20}{l}}
  {{\cal{P}}_2}:\mathop {\min }\limits_{{{\mathbf{t}}_k}} \;\;{\left\| {\sum\limits_{k = 1}^K {{{\mathbf{t}}_k}e^{ {j\left({\phi _k}-{\phi_1}\right)} }} } \right\|^2}  \hfill \\
  \displaystyle s.t.{\kern 1pt} {\kern 1pt} {\kern 1pt} {\kern 1pt} {\kern 1pt} \left| {\operatorname{Im} \left( {{\mathbf{h}}_i^T\sum\limits_{k = 1}^K {{{\mathbf{t}}_k}e^{j\left( {{\phi _k} - {\phi _i}} \right)}} } \right)} \right| \le \left( {\operatorname{Re} \left( {{\mathbf{h}}_i^T\sum\limits_{k = 1}^K {{{\mathbf{t}}_k}e^{j\left( {{\phi _k} - {\phi _i}} \right)} } } \right) - \sqrt {{\Gamma _i}\left( {\sigma _C^2 + {P_R}{{\left\| {{{\mathbf{f}}_i}} \right\|}^2}} \right)} } \right)\tan  {\psi } ,\forall i, \hfill \\
  {\kern 1pt} {\kern 1pt} {\kern 1pt} {\kern 1pt} {\kern 1pt}{\kern 1pt} {\kern 1pt} {\kern 1pt} {\kern 1pt} {\kern 1pt}{\kern 1pt} {\kern 1pt} {\kern 1pt} {\kern 1pt} {\kern 1pt}{\kern 1pt} {\kern 1pt} {\kern 1pt} {\kern 1pt} {\kern 1pt}{\kern 1pt}{\left| {{\mathbf{g}}_m^T\sum\limits_{k = 1}^K {{{\mathbf{t}}_k}} {e^{j\phi_k}}} \right|^2} \le {R_m}\sigma _R^2,\forall m. 
\end{array} 
\end{equation}
\setcounter{equation}{\value{MYtempeqncnt}}
\hrulefill
\vspace*{4pt}
\end{figure*}
Following the virtual multicast model in \cite{7103338}, the power minimization problem ${{\cal{P}}_2}$ can be equivalently written as
\setcounter{equation}{11}
\begin{equation}
\begin{array}{*{20}{l}}
  {{\cal{P}}_3}:\mathop {\min }\limits_{\mathbf{w}}  {\left\| {\mathbf{w}} \right\|^2} \hfill \\
 \displaystyle s.t.{\kern 1pt}\left| {\operatorname{Im} \left( {{\mathbf{{\tilde h}}}_i^T{\mathbf{w}}} \right)} \right| \le \left( {\operatorname{Re} \left( {{\mathbf{{\tilde h}}}_i^T{\mathbf{w}}} \right) - \sqrt {{{\tilde \Gamma }_i}} } \right)\tan {\psi }, \forall i, \hfill \\
  \;\;\;\;\;{\kern 1pt}\left| {{\mathbf{\tilde g}}_m^T{\mathbf{w}}} \right| \le \sqrt {{R_m}\sigma _R^2} ,{\kern 1pt} {\kern 1pt} \forall m, \hfill \\ 
\end{array}
\end{equation}
where $ {\mathbf{w}} \triangleq \sum\limits_{k = 1}^K {{{\mathbf{t}}_k}{e^{j\left( {{\phi _k} - {\phi _1}} \right)}}}$, ${{\mathbf{{\tilde h}}}_i} \triangleq {{\mathbf{h}}_i}{e^{j\left( {{\phi _1} - {\phi _i}} \right)}}$, ${{\mathbf{\tilde g}}_m} \triangleq {{\mathbf{g}}_m}{e^{j{\phi _1}}}$, ${{\tilde \Gamma }_i}={\Gamma _i}\left( {\sigma _C^2 + {P_R}{{\left\| {{{\mathbf{f}}_i}} \right\|}^2}} \right)$.
Similarly, the CI-based interference minimization problem is given by
\begin{equation}
\begin{array}{*{20}{l}}
  {{\cal{P}}_4}:\mathop {\min }\limits_{\mathbf{w}} \sum\limits_{m = 1}^M {{{\left| {{\mathbf{\tilde g}}_m^T{\mathbf{w}}} \right|}^2}}  \hfill \\
 \displaystyle s.t.{\kern 1pt}\left| {\operatorname{Im} \left( {{\mathbf{{\tilde h}}}_i^T{\mathbf{w}}} \right)} \right| \le \left( {\operatorname{Re} \left( {{\mathbf{{\tilde h}}}_i^T{\mathbf{w}}} \right) - \sqrt {{{\tilde \Gamma }_i}} } \right)\tan{\psi }, \forall i, \hfill \\
  \;\;\;\;\;{\kern 1pt}\;{\left\| {\mathbf{w}} \right\|} \le \sqrt{P}.
\end{array}
\end{equation}
After obtaining the optimal solution ${\bf{w}}$, the beamforming vectors can be obtained as
\begin{equation}
{{\bf t}_1} = \frac{{\mathbf{w}}}{K},
\end{equation}
\begin{equation}
{{\bf t}_k} = \frac{{{\mathbf{w}}{e^{j\left( {{\phi _1} - {\phi _k}} \right)}}}}{K},\forall k.
\end{equation}
Note that both ${{\cal{P}}_3}$ and ${{\cal{P}}_4}$ are convex and can be easily solved by numerical tools. To make the proposed method more realizable in practical scenarios, we will take ${{\cal{P}}_3}$ as an example to derive an efficient algorithm to solve it, and a similar algorithm can be also applied to ${{\cal{P}}_4}$.

\section{Efficient Algorithm for Power Minimization Beamforming} 
\subsection{Real representation of the problem}
For the ease of our further analysis, we first derive the real representation of the problem. Let us rewrite the related channel vectors and the beamforming vector as follows
\begin{equation}
\begin{array}{*{20}{l}}
  {{{\mathbf{{\tilde h}}}}_i} = {{{\mathbf{{\tilde h}}}}_{Ri}} + j{{{\mathbf{{\tilde h}}}}_{Ii}}, \hfill \\
  {{{\mathbf{\tilde g}}}_m} = {{{\mathbf{\tilde g}}}_{Rm}} + j{{{\mathbf{\tilde g}}}_{Im}}, \hfill \\
  {\mathbf{w}} = {{\mathbf{w}}_R} + j{{\mathbf{w}}_I}, \hfill \\ 
\end{array}
\end{equation}
where 
\begin{equation}
\begin{array}{*{20}{l}}
  {{{\mathbf{{\tilde h}}}}_{Ri}} = \operatorname{Re} \left( {{{{\mathbf{{\tilde h}}}}_i}} \right),{{{\mathbf{{\tilde h}}}}_{Ii}} = \operatorname{Im} \left( {{{{\mathbf{{\tilde h}}}}_i}} \right), \hfill \\
  {{{\mathbf{\tilde g}}}_{Rm}} = \operatorname{Re} \left( {{{{\mathbf{\tilde g}}}_m}} \right),{{{\mathbf{\tilde g}}}_{Im}} = \operatorname{Im} \left( {{{{\mathbf{\tilde g}}}_m}} \right), \hfill \\
  {{\mathbf{w}}_R} = \operatorname{Re} \left( {\mathbf{w}} \right),{{\mathbf{w}}_I} = \operatorname{Im} \left( {\mathbf{w}} \right). \hfill \\ 
\end{array}
\end{equation}
Then we define the following real-valued vectors and matrices
\begin{equation}
\begin{array}{*{20}{l}}
  {{{\mathbf{{\bar h}}}}_i} = \left[ {{{{\mathbf{{\tilde h}}}}_{Ri}};{{{\mathbf{{\tilde h}}}}_{Ii}}} \right], \hfill \\
  {{\mathbf{w}}_1} = \left[ {{{\mathbf{w}}_I};{{\mathbf{w}}_R}} \right],{{\mathbf{w}}_2} = \left[ {{{\mathbf{w}}_R}; - {{\mathbf{w}}_I}} \right], \hfill \\
  {{\pmb{\beta}}_m}{\text{ = }}\left[ {\begin{array}{*{20}{c}}
  {{{{\mathbf{\tilde g}}}_{Rm}}}&{{{{\mathbf{\tilde g}}}_{Im}}} \\ 
  {{{{\mathbf{\tilde g}}}_{Im}}}&{ - {{{\mathbf{\tilde g}}}_{Rm}}} 
\end{array}} \right],{\mathbf{\Pi }} = \left[ {\begin{array}{*{20}{c}}
  {\mathbf{0}}_K&{{\text{ - }}{\mathbf{I}}_K} \\ 
  {\mathbf{I}}_K&{\mathbf{0}}_K 
\end{array}} \right],
\end{array}
\end{equation}
where ${\bf{I}}_K$ and ${\bf{0}}_K$ denote the $K\times K$ identity matrix and all-zero matrix respectively. Thus we obtain
\begin{equation}
\begin{array}{*{20}{l}}
  \operatorname{Re} \left( {{\mathbf{{\tilde h}}}_i^T{\mathbf{w}}} \right) = {{{\mathbf{{\bar h}}}}_i}^T{{\mathbf{w}}_2}, \hfill \\
  \operatorname{Im} \left( {{\mathbf{{\tilde h}}}_i^T{\mathbf{w}}} \right) = {{{\mathbf{{\bar h}}}}_i}^T{{\mathbf{w}}_1} = {{{\mathbf{{\bar h}}}}_i}^T{\mathbf{\Pi }}{{\mathbf{w}}_2} \triangleq {\mathbf{b}}_i^T{{\mathbf{w}}_2}, \hfill \\
  {\left| {{\mathbf{\tilde g}}_m^T{\mathbf{w}}} \right|^2} = {\left\| {\left[ {\begin{array}{*{20}{c}}
  {{\mathbf{\tilde g}}_{Rm}^T}&{{\mathbf{\tilde g}}_{Im}^T} \\ 
  {{\mathbf{\tilde g}}_{Im}^T}&{ - {\mathbf{\tilde g}}_{Rm}^T} 
\end{array}} \right]\left[ {\begin{array}{*{20}{c}}
  {{{\mathbf{w}}_R}} \\ 
  { - {{\mathbf{w}}_I}} 
\end{array}} \right]} \right\|^2} = {\left\| {{\pmb{\beta }}_m^T{{\mathbf{w}}_2}} \right\|^2}. \hfill \\ 
\end{array}
\end{equation}
Finally, the real version of the problem is given as
\begin{equation}
\begin{array}{*{20}{l}}
  {{\cal{P}}_5}: \mathop {\min }\limits_{{\mathbf{w}}_2} \;\;{\kern 1pt} {\kern 1pt} {\left\| {{{\mathbf{w}}_2}} \right\|^2} \hfill \\
  \displaystyle s.t.{\kern 1pt} {\kern 1pt} {\kern 1pt} {\kern 1pt} {\kern 1pt} {\kern 1pt} {\kern 1pt} {\kern 1pt} {\mathbf{b}}_i^T{{\mathbf{w}}_2} - {{{\mathbf{{\bar h}}}}_i}^T{{\mathbf{w}}_2}\tan {\psi } + \sqrt {{{\tilde \Gamma }_i}} \tan {\psi } \le 0,\forall i, \hfill \\
  \displaystyle \;\;\;\;\;\; - {\mathbf{b}}_i^T{{\mathbf{w}}_2} - {{{\mathbf{{\bar h}}}}_i}^T{{\mathbf{w}}_2}\tan{\psi } + \sqrt {{{\tilde \Gamma }_i}} \tan {\psi }\le 0,\forall i, \hfill \\
  \;\;\;\;\;\;{\left\| {{\pmb{\beta }}_m^T{{\mathbf{w}}_2}} \right\|^2} \le {R_m}\sigma _R^2,\forall m. 
\end{array}
\end{equation}
\subsection{The Dual Problem}
In order to reveal the structure of the solution, we formulate the dual problem of ${{\cal{P}}_5}$. Let us define the dual variable that associate with the three constraints in (20) as $ {\mathbf{u}},{\mathbf{v}},{\mathbf{c}}$ respectively, where $u_i \ge 0, v_i \ge 0, c_m \ge 0, \forall i, \forall m$ are the elements of the three dual vectors. The corresponding Lagrangian is given as (21) at the top of the next page. 
\begin{figure*}[!t]
\normalsize
\newcounter{MYtempeqncnt1}
\setcounter{MYtempeqncnt1}{\value{equation}}
\setcounter{equation}{20}
\begin{small}
\begin{equation}
\begin{array}{*{20}{l}}
 \displaystyle \mathcal{L}\left( {{{\mathbf{w}}_2},{\mathbf{u}},{\mathbf{v}},{\mathbf{c}}} \right) \hfill \\
 \displaystyle = {\left\| {{{\mathbf{w}}_2}} \right\|^2} + \sum\limits_{i = 1}^K {{u_i}\left( {{\mathbf{b}}_i^T{{\mathbf{w}}_2} - {\mathbf{{\bar h}}}_i^T{{\mathbf{w}}_2}\tan \psi  + \sqrt {{{\tilde \Gamma }_i}} \tan \psi } \right)}  \hfill \\
 \displaystyle + \sum\limits_{i = 1}^K {{v_i}\left( { - {\mathbf{b}}_i^T{{\mathbf{w}}_2} - {\mathbf{{\bar h}}}_i^T{{\mathbf{w}}_2}\tan \psi  + \sqrt {{{\tilde \Gamma }_i}} \tan \psi } \right)}  + \sum\limits_{m = 1}^M {{c_m}\left( {{{\left\| {\pmb \beta _m^T{{\mathbf{w}}_2}} \right\|}^2} - {R_m}\sigma _R^2} \right)}  \hfill \\
 \displaystyle = {\mathbf{w}}_2^T\left( {{\bf I} + \sum\limits_{m = 1}^M {{c_m}}\pmb \beta _m \pmb \beta _m^T} \right){{\mathbf{w}}_2} + \sum\limits_{i = 1}^K {{\kern 1pt} \left[ {\left( {{u_i} - {v_i}} \right){\mathbf{b}}_i^T - \left( {{u_i} + {v_i}} \right){\mathbf{{\bar h}}}_i^T\tan \psi } \right]} {{\mathbf{w}}_2} 
 \displaystyle +  \tan \psi \sum\limits_{i = 1}^K {\sqrt {{{\tilde \Gamma }_i}} \left( {{u_i} + {v_i}} \right)}  - {R_m}\sigma _R^2\sum\limits_{m = 1}^M {{c_m}}.
\end{array} 
\end{equation}
\end{small}
\setcounter{equation}{\value{MYtempeqncnt1}}
\hrulefill
\vspace*{4pt}
\end{figure*}

By the following definitions
\setcounter{equation}{21}
\begin{equation}
\begin{array}{*{20}{l}}
{\mathbf{{\bar h}}} = \left[ {{{{\mathbf{{\bar h}}}}_1},{{{\mathbf{{\bar h}}}}_2},...,{{{\mathbf{{\bar h}}}}_K}} \right],
{\mathbf{b}} = \left[ {{{\mathbf{b}}_1},{{\mathbf{b}}_2},...,{{\mathbf{b}}_K}} \right], {\mathbf{1}} = \left[ {{{\mathbf{I}}_K};{{\mathbf{I}}_K}} \right], \hfill \\
{\pmb{\lambda }} = \left[ {{\mathbf{u}};{\mathbf{v}}} \right],  {\pmb{\beta }} = \left[ {{{\pmb{\beta }}_1},{{\pmb{\beta }}_2},...,{{\pmb{\beta }}_M}} \right], {\mathbf{R}} = \left[ {{R_1},{R_2},...,{R_M}} \right],\hfill \\
{\mathbf{c}} = \left[ {{c_1};{c_2};...;{c_{{M}}}} \right], {\mathbf{\tilde c}} = \left[ {{c_1};{c_1};{c_2};{c_2};...;{c_{{M}}};{c_{{M}}}} \right], \hfill \\
\displaystyle {\mathbf{\tilde \Gamma }} = \left[ {{\tilde\Gamma _1};{\tilde\Gamma _2};...;{\tilde\Gamma _K}} \right], {\mathbf{A}} = \left[ {{\mathbf{{\bar h}}}\tan \psi  - {\mathbf{b}},{\mathbf{{\bar h}}}\tan \psi  + {\mathbf{b}}} \right], \hfill \\
\end{array}
\end{equation}
the Lagrangian can be further simplified as
\begin{equation}
\begin{gathered}
  \mathcal{L}\left( {{{\mathbf{w}}_2},{\mathbf{u}},{\mathbf{v}},{\mathbf{c}}} \right) \hfill \\
   = {\mathbf{w}}_2^T\left( {{\mathbf{I}} + {\pmb{\beta }}\operatorname{diag} \left( {{\mathbf{\tilde c}}} \right){{\pmb{\beta }}^T}} \right){{\mathbf{w}}_2} + {{\pmb{\lambda }}^T}{{\mathbf{A}}^T}{{\mathbf{w}}_2} \hfill \\
   + \tan \psi  \sqrt {{{{\mathbf{\tilde \Gamma }}}^T}} {{\mathbf{1}}^T}{\pmb{\lambda }} - \sigma _R^2{{\mathbf{R}}^T}{\mathbf{c}}, \hfill \\ 
\end{gathered}
\end{equation}
where $\operatorname{diag}(\bf{x})$ denotes the diagonal matrix whose diagonal elements are given by $\bf{x}$. Let $ \frac{{\partial \mathcal{L}}}{{\partial {{\mathbf{w}}_2}}} = 0$, the optimal solution of $\bf w_2$ is given by
\begin{equation}
{\mathbf{w}}_2^* =  - \frac{{{{\left( {{\mathbf{I}} + {\pmb{\beta }}\operatorname{diag} \left( {{\mathbf{\tilde c}}} \right){{\pmb{\beta }}^T}} \right)}^{ - 1}}{\mathbf{A \pmb \lambda }}}}{2},
\end{equation}
which implies ${\pmb{\lambda}}  \ne \bf{0} $, for the reason that ${\pmb{\lambda}}  = \bf{0} $ yields the trivial solution of ${\mathbf{w}}_2^*=\bf{0}$. Substituting the optimal ${\mathbf{w}}_2^*$ into the Lagrangian leads to
\begin{equation}
\begin{gathered}
  \mathcal{L}\left( {{\mathbf{u}},{\mathbf{v}},{\mathbf{c}}} \right) =  - \frac{1}{4}{{\pmb{\lambda  }}^T}{{\mathbf{A}}^T}{\left( {{\mathbf{I}}{\text{ + }}{\pmb{\beta}}\operatorname{diag} \left( {{\mathbf{\tilde c}}} \right){{\pmb{\beta}}^T}} \right)^{ - 1}}{\mathbf{A\pmb\lambda }} \hfill \\
   \quad + \tan \psi \sqrt {{{{\mathbf{\tilde \Gamma }}}^T}} {{\mathbf{1}}^T}{\pmb{\lambda  }} - \sigma _R^2{{\mathbf{R}}^T}{\mathbf{c}}. \hfill \\ 
\end{gathered}
\end{equation}
Therefore, the dual problem is given as
\begin{equation}
\begin{gathered}
  {{\cal{P}}_6}:\;\;\mathop{\max} \limits_{\pmb{\lambda}, \mathbf{c}} \;\; - \frac{1}{4}{{\pmb{\lambda  }}^T}{{\mathbf{A}}^T}{\left( {{\mathbf{I}}{\text{ + }}{\pmb{\beta}}\operatorname{diag} \left( {{\mathbf{\tilde c}}} \right){{\pmb{\beta}}^T}} \right)^{ - 1}}{\mathbf{A\pmb\lambda }} \hfill \\
  \;\;\;\;\;\;\;\;\;\;\;\;\;\; \quad \quad + \tan \psi \sqrt {{{{\mathbf{\tilde \Gamma }}}^T}} {{\mathbf{1}}^T}{\pmb{\lambda  }} - \sigma _R^2{{\mathbf{R}}^T}{\mathbf{c}} \hfill \\
  s.t.\;\;\;{\pmb{\lambda  }} \ge {\mathbf{0}},{\mathbf{c}} \ge {\mathbf{0}}. \hfill \\ 
\end{gathered}
\end{equation}
Note that when removing the INR constraints, the dual problem is the same as the original CI-based power minimization problem in \cite{7103338}.
\subsection{Karush-Kuhn-Tucker Conditions}
Let us first rewrite the dual problem as the following standard convex form
\begin{equation}
\begin{gathered}
  {{\cal{P}}_7}:\;\;\mathop{\min} \limits_{\pmb{\lambda}, \mathbf{c}} \;\;f\left( {{\pmb{\lambda }},{\mathbf{c}}} \right)= \frac{1}{4}{{\pmb{\lambda  }}^T}{{\mathbf{A}}^T}{\left( {{\mathbf{I}}{\text{ + }}{\pmb{\beta}}\operatorname{diag} \left( {{\mathbf{\tilde c}}} \right){{\pmb{\beta}}^T}} \right)^{ - 1}}{\mathbf{A\pmb\lambda }} \hfill \\
  \;\;\;\;\;\;\;\;\;\;\;\;\;\; \quad \quad - \tan \psi \sqrt {{{{\mathbf{\tilde \Gamma }}}^T}} {{\mathbf{1}}^T}{\pmb{\lambda  }} + \sigma _R^2{{\mathbf{R}}^T}{\mathbf{c}} \hfill \\
  s.t.\;\;\;{\pmb{\lambda  }} \ge {\mathbf{0}},{\mathbf{c}} \ge {\mathbf{0}}. \hfill \\ 
\end{gathered}
\end{equation}
It is easy to observe that the primal problem ${{\cal{P}}_5}$ is a standard Quadratically Constrained Quadratic Program (QCQP), and is convex and the strong duality holds, thus the Karush-Kuhn-Tucker (KKT) Conditions are sufficient for primal and dual optimal variables \cite{boyd2004convex}, which are denoted by ${\mathbf{w}}_2^*$, $\pmb \lambda^*$ and $\bf c^*$ respectively, and they have zero duality gap. Based on the complementary slackness conditions we have
\begin{equation}
\begin{gathered}
  c_m^*\left( {{\left\| {\pmb \beta _m^T{{\mathbf{w}}_2}} \right\|}^2 - {R_m}\sigma _R^2} \right) = 0,\forall m.\hfill \\
\end{gathered}
\end{equation}
When removing the INR constraints, the optimization ${{\cal{P}}_5}$ has the same structure with the original CI-based power minimization problem\cite{7103338}, which is given as
\begin{equation}
\begin{array}{*{20}{l}}
  {{\cal{P}}_8}:{\kern 1pt} \;\;\mathop {\min }\limits_{{\mathbf{w}}_2} \;\;{\kern 1pt} {\kern 1pt} {\left\| {{{\mathbf{w}}_2}} \right\|^2} \hfill \\
  \displaystyle s.t.{\kern 1pt} {\kern 1pt} {\kern 1pt} {\kern 1pt} {\kern 1pt} {\kern 1pt} {\kern 1pt} {\kern 1pt} {\mathbf{b}}_i^T{{\mathbf{w}}_2} - {{{\mathbf{{\bar h}}}}_i}^T{{\mathbf{w}}_2}\tan {\psi } + \sqrt {{{\tilde \Gamma }_i}} \tan {\psi } \le 0,\forall i, \hfill \\
  \displaystyle \;\;\;\;\;\; - {\mathbf{b}}_i^T{{\mathbf{w}}_2} - {{{\mathbf{{\bar h}}}}_i}^T{{\mathbf{w}}_2}\tan{\psi } + \sqrt {{{\tilde \Gamma }_i}} \tan {\psi }\le 0,\forall i. \hfill \\
\end{array}
\end{equation}
The dual problem of ${{\cal{P}}_8}$ is given by
\begin{equation}
\begin{gathered}
  {{\cal{P}}_9}:\mathop{\min} \limits_{\pmb{\lambda}} \;\;\frac{{{{\left\| {{\mathbf{A\pmb\lambda }}} \right\|}^2}}}{4} - \tan \psi \sqrt {{{{\mathbf{\tilde \Gamma }}}^T}} {{\mathbf{1}}^T}{\pmb{\lambda }} \hfill \\
  s.t.\;\;{\pmb{\lambda }} \ge 0, \hfill \\ 
\end{gathered}
\end{equation}
and the optimal solution to ${\cal P}_8$ has the structure of
\begin{equation}
{\mathbf{w}}_2^* =  - \frac{{{\mathbf{A\pmb\lambda }}}}{2},
\end{equation}
where $\bf A$ and $\pmb \lambda$ are defined in (22). By substituting (31) in the INR constraint to obtain
\begin{equation}
{\left\| {{\pmb{\beta }}_m^T{{\mathbf{w}}_2}} \right\|^2}= \frac{1}{4}{\left\| {{\pmb{\beta }}_m^T{\mathbf{A \pmb\lambda }}} \right\|^2}.
\end{equation}
Therefore, if ${R_m}\sigma _R^2\ge\frac{1}{4}{\left\| {{\pmb{\beta }}_m^T{\mathbf{A \pmb\lambda }}} \right\|^2}$, (31) is a feasible point for ${{\cal{P}}_5}$. Since (31) is the optimal point of ${{\cal{P}}_8}$, this implies that it is also the optimal point for ${{\cal{P}}_5}$ for the reason that the minimum value of ${{\cal{P}}_5}$ will always be greater than or equal to the minimum value of ${{\cal{P}}_8}$ due to the extra INR constraint. Thus the related INR constraint will always be satisfied, and $c_m=0$. By denoting the optimal solution of ${{\cal{P}}_9}$ by $\pmb \lambda_0$, the following corollary holds immediately.
\\\indent \emph{Corollary 1:} If ${R_m}\sigma _R^2>\frac{1}{4}{\left\| {{\pmb{\beta }}_m^T{\mathbf{A \pmb\lambda_0 }}} \right\|^2}$,  ${{\cal{P}}_5}$ is equivalent to the original CI problem ${{\cal{P}}_8}$, where $\pmb\lambda_0$ is the optimal solution of ${{\cal{P}}_9}$.

\subsection{Efficient Gradient Projection Method}
The closed form of the optimal solution to ${{\cal{P}}_7}$ is difficult to derive. Nevertheless, thanks to the simple constraints with only bounds on the variables, it is convenient to apply a gradient projection algorithm to solve the problem\cite{wright1999numerical}. Let us first derive the gradient of the dual function as follows. By letting ${\mathbf{M}} = {\left( {{\mathbf{I}}{\text{ + }}{\pmb{\beta }}\operatorname{diag} \left( {{\mathbf{\tilde c}}} \right){{\pmb{\beta }}^T}} \right)^{ - 1}}$, the derivative is given as
\begin{equation}
\begin{gathered}
  \frac{{\partial f}}{{\partial {\pmb{\lambda }}}} = \frac{1}{2}{{\pmb{\lambda }}^T}{{\mathbf{A}}^T}{\mathbf{MA}} - \tan \psi \sqrt {{{{\mathbf{\tilde \Gamma }}}^T}} {{\mathbf{1}}^T}, \hfill \\
  \frac{{\partial f}}{{\partial {c_m}}} =  - \frac{1}{4}{\left| {{{\pmb{\lambda }}^T}{{\mathbf{A}}^T}{\mathbf{M}}{{\pmb{\beta }}_m}} \right|^2} + \sigma _R^2{R_m},\forall m. \hfill \\ 
\end{gathered}
\end{equation}
Thus the gradient is give by
\begin{equation}
\begin{gathered}
  \triangledown f\left( {{\pmb{\lambda }},{\mathbf{c}}} \right) = {\left[ {\frac{{\partial f}}{{\partial {\pmb{\lambda }}}},\frac{{\partial f}}{{\partial {\mathbf{c}}}}} \right]^T} \hfill \\
   \quad = \left[ \begin{gathered}
  \frac{1}{2}{{\mathbf{A}}^T}{\mathbf{MA \pmb\lambda }} - \tan \psi {{\mathbf{1}}} \sqrt {{{{\mathbf{\tilde \Gamma }}}}} ; \hfill \\
   - \frac{1}{4}{\left| {{{\pmb{\lambda }}^T}{{\mathbf{A}}^T}{\mathbf{M}}{{\pmb{\beta }}_1}} \right|^2} + \sigma _R^2{R_1}; \hfill \\
   - \frac{1}{4}{\left| {{{\pmb{\lambda }}^T}{{\mathbf{A}}^T}{\mathbf{M}}{{\pmb{\beta }}_2}} \right|^2} + \sigma _R^2{R_2}; \hfill \\
  ... \hfill \\
   - \frac{1}{4}{\left| {{{\pmb{\lambda }}^T}{{\mathbf{A}}^T}{\mathbf{M}}{{\pmb{\beta }}_M}} \right|^2} + \sigma _R^2{R_M} \hfill \\ 
\end{gathered}  \right]. \hfill \\ 
\end{gathered}
\end{equation}
Based on above derivations, the following Algorithm 1 is proposed to solve problem ${{\cal{P}}_7}$, where we use an iterative gradient projection method, and the step size can be decided by the Armijo rule or other backtracking linesearch methods \cite{wright1999numerical}. After obtaining the optimal ${\bf w}_2$, the beamforming vectors can be calculated by (14) and (15). 
\renewcommand{\algorithmicrequire}{\textbf{Input:}}
\renewcommand{\algorithmicensure}{\textbf{Output:}}
\begin{figure}[htbp]
  \rule{\linewidth}{.1pt}
  {\bf{Algorithm 1}}
  \begin{algorithmic}[1]
    \REQUIRE ${\mathbf{H}},{\mathbf{G}},{\mathbf{F}},{\mathbf{\Gamma }},{\mathbf{R}},{\sigma _c},{\sigma _R}$.
    \ENSURE Optimal solution ${{\bf w}^*_2}$ for problem ${{\cal{P}}_5}$. \\
    \STATE Initialize randomly $\pmb \lambda^{(0)} \ge 0, \mathbf c^{(0)} \ge 0 $.
    \STATE In the \emph{i}th iteration, update $\pmb \lambda$ and $\bf c$ by:
    \begin{small}
    \begin{equation*}
        \left[ {{{\pmb{\lambda }}^{\left(i \right)}},{{\mathbf{c}}^{\left( i \right)}}} \right] = \max \left( {\left[ {{{\pmb{\lambda }}^{\left( i \right)}},{{\mathbf{c}}^{\left( i \right)}}} \right] - {a_i}\triangledown f\left( {{{\pmb{\lambda }}^{\left( {i - 1} \right)}},{{\mathbf{c}}^{\left( {i - 1} \right)}}} \right),{\mathbf{0}}} \right),
    \end{equation*}
    \end{small}
where the step size $a_i$ is calculated by the backtracking linesearch method.
    \STATE Go back to 2 until convergence. 
    \STATE Calculate ${{\bf w}^*_2}$ by \hfill
    \begin{small}
    \begin{equation*}
        {\mathbf{w}}_2^* =  - \frac{{{{\left( {{\mathbf{I}} + {\pmb{\beta }}\operatorname{diag} \left( {{\mathbf{\tilde c}^{(i)}}} \right){{\pmb{\beta }}^T}} \right)}^{ - 1}}{\mathbf{A \pmb \lambda }^{(i)}}}}{2}.
    \end{equation*}
    \end{small}
    \STATE \textbf{end}
  \end{algorithmic}
  \rule{\linewidth}{.1pt}
\end{figure}

\section{Impact on Radar Performance}
\subsection{SDR-based beamforming}
The interference from BS to radar will have an impact on radar's performance, which will lower the detection probability and the accuracy for Direction of Arrival (DoA) estimation. First we consider the detection problem. Note that the target detection process can be described as a Binary Hypothesis Testing problem, which is given by

\begin{equation}
{{\bf{y}}^R_l}=\left\{ \begin{gathered}
  {{\cal H}_1}:\alpha \sqrt{P_R} {\bf{A}}\left( \theta  \right){{\bf s}_l} + {{\bf{G}}^T}\sum\limits_{k = 1}^K {{{\bf{t}}_k}{d_k}\left[ l \right]} + {\mathbf{z}}_l,\hfill \\\;\;\;\;\;\;\;\; l = 1,2,...,L, \hfill \\
  {{\cal H}_0}:{{\bf{G}}^T}\sum\limits_{k = 1}^K {{{\bf{t}}_k}{d_k}\left[ l \right]} + {\mathbf{z}}_l,\;\;l = 1,2,...,L. \hfill \\ 
\end{gathered}  \right.
\end{equation}
\\\indent Due to the unknown parameters $\alpha$ and $\theta$, we use the Generalized Likelihood Ratio Test (GLRT) method to solve the above problem. Consider the sufficient statistic of the received signal, which is obtained by matched filtering \cite{1703855}, and is given by
\begin{equation}
\begin{gathered}
  {\mathbf{\tilde Y}} = \frac{1}{{\sqrt L }}\sum\limits_{l = 1}^L {{\bf{y}}^R_l}{{\mathbf{s}}^H_l} \hfill \\
   \quad= \alpha {\sqrt {LP_R} }{\mathbf{A}}\left( \theta  \right) 
   + \frac{1}{{\sqrt L }}\sum\limits_{l = 1}^L {\left( {{{\mathbf{G}}^T}\sum\limits_{k = 1}^K {{{\mathbf{t}}_k}{d_k}\left[ l \right] + {\mathbf{z}}_l} } \right)} {{\mathbf{s}}^H_l} \hfill \\ 
\end{gathered}
\end{equation}
Let ${\mathbf{\tilde y}}$ be the vectorization of ${\mathbf{\tilde Y}}$, we have
\begin{equation}
\begin{gathered}
  {\mathbf{\tilde y}} = \operatorname{vec} \left( {{\mathbf{\tilde Y}}} \right) \hfill \\
   \quad= \alpha {\sqrt {LP_R} }\operatorname{vec} \left( {{\mathbf{A}}\left( \theta  \right)} \right) \hfill \\
   \quad + \operatorname{vec} \left( \frac{1}{{\sqrt L }}\sum\limits_{l = 1}^L {\left( {{{\mathbf{G}}^T}\sum\limits_{k = 1}^K {{{\mathbf{t}}_k}{d_k}\left[ l \right] + {\mathbf{z}}_l} } \right)} {{\mathbf{s}}^H_l} \right) \hfill \\
   \quad\triangleq \alpha {\sqrt {LP} _R}\operatorname{vec} \left( {{\mathbf{A}}\left( \theta  \right)} \right) + {\pmb{\varepsilon }}, \hfill \\ 
\end{gathered}
\end{equation}
where $\pmb \varepsilon$ is zero-mean, complex Gaussian distributed, and has the block covariance matrix as
\begin{equation}
{\mathbf{C}} = \left[ {\begin{array}{*{20}{c}}
  {{\mathbf{J}} + \sigma _R^2{{\mathbf{I}}_M}}&{}&{\mathbf{0}} \\ 
  {}&{...}&{} \\ 
  {\mathbf{0}}&{}&{{\mathbf{J}} + \sigma _R^2{{\mathbf{I}}_M}}
\end{array}} \right],
\end{equation}
where ${\mathbf{C}} \in {\mathbb{C}}^{{M^2}\times {M^2}}$, and ${\mathbf{J}}={{\mathbf{G}}^T}\sum\limits_{k = 1}^K {{{\mathbf{t}}_k}{\mathbf{t}}_k^H} {{\mathbf{G}}^*} $.
Hence, (35) is equivalent to the following hypothesis:
\begin{equation}
{\mathbf{\tilde y}} = \left\{ \begin{gathered}
  {{\cal H}_1}:\alpha {\sqrt {LP_R} }{\bf d}(\theta) + {\mathbf{\pmb\varepsilon }}, \hfill \\
  {{\cal H}_0}:{\mathbf{\pmb\varepsilon }}, \hfill \\ 
\end{gathered}  \right.
\end{equation}
where ${\bf d}(\theta) = \operatorname{vec} \left( {{\mathbf{A}}\left( \theta  \right)} \right)$. As per the standard GLRT decision rule, if 
\begin{equation}
{L_{\tilde{\bf y}}\left({\hat \alpha}, {\hat \theta}\right)}=\frac{{p\left( {{\mathbf{\tilde y}};{\hat \alpha}, {\hat \theta}, {{\cal H}_1}} \right)}}{{p\left( {{\mathbf{\tilde y}}; {{\cal H}_0}} \right)}} > \eta, 
\end{equation}
then ${\cal H}_1$ is chosen, where ${p\left( {{\mathbf{\tilde y}};{\hat \alpha}, {\hat \theta}, {{\cal H}_1}} \right)}$ and ${p\left( {{\mathbf{\tilde y}};{{\cal H}_0}} \right)}$ are the Probability Density Function (PDF) under ${\cal{H}}_1$ and ${\cal{H}}_0$ respectively, $\hat \alpha$ and $\hat \theta$ is the maximum likelihood estimation (MLE) of $\alpha$ and $\theta$ under ${\cal{H}}_1$, and is given by $\left[ {\hat \alpha ,\hat \theta } \right] = \mathop {\max }\limits_{\alpha ,\theta } p\left( {{\mathbf{\tilde y}}\left| {\alpha ,\theta ,{\cal H}_1} \right.} \right)$, $\eta$ is the decision threshold. (39) can be viewed as a hypothesis testing problem for MIMO radar detection in the homogeneous Gaussian clutter with covariance matrix $\bf C$, which has been discussed in\cite{5545189}. In this case, the GLRT detection statistic is given by
\begin{equation}
\begin{array}{*{20}{l}}
\displaystyle \ln {L_{{\mathbf{\tilde y}}}}\left( {\hat \theta } \right) = \frac{{{{\left| {{{\mathbf{d}}^H}\left( {\hat \theta } \right){{\mathbf{C}}^{ - 1}}{\mathbf{\tilde y}}} \right|}^2}}}{{{{\mathbf{d}}^H}\left( {\hat \theta } \right){{\mathbf{C}}^{ - 1}}{\mathbf{d}}\left( {\hat \theta } \right)}}\hfill \\
\displaystyle  \;\;\;\;\;\;\;\;\;\;\;\;\;\;\;\;= \frac{{{{\left| {{\text{tr}}\left( {{\mathbf{\tilde Y}}{{\mathbf{A}}^H}\left( {\hat \theta } \right){{{\mathbf{\tilde J}}}^{ - 1}}} \right)} \right|}^2}}}{{{\text{tr}}\left( {{\mathbf{A}}\left( {\hat \theta } \right){{\mathbf{A}}^H}\left( {\hat \theta } \right){{{\mathbf{\tilde J}}}^{ - 1}}} \right)}}\mathop {\gtrless}\limits_{{\cal H}_0}^{{\cal H}_1} \eta,
\end{array}
\end{equation}
where ${{\mathbf{\tilde J}}}={{\mathbf{J}} + \sigma _R^2{{\mathbf{I}}_M}}$. According to \cite{kay1998fundamentals2}, the asymptotic distribution of (41) is given by
\begin{equation}
\ln {L_{{\mathbf{\tilde y}}}}\left( {\hat \theta } \right) \sim \left\{ \begin{gathered}
  {{\cal H}_1}:{\mathcal X}_2^2\left( \rho  \right), \hfill \\
  {{\cal H}_0}:{\mathcal X}_2^2, \hfill \\ 
\end{gathered}  \right.
\end{equation}
where $ {\mathcal X}_2^2 $ and $ {\mathcal X}_2^2\left( \rho  \right) $ are central and non-central chi-squared distributions with two Degrees of Freedom (DoFs), and $\rho$ is the non-central parameter, which is given by
\begin{equation}
\begin{gathered}
  \rho  = {|\alpha| ^2}L{P_R}{\operatorname{vec} ^H}\left( {{\mathbf{A}}\left( \theta  \right)} \right){{\mathbf{C}}^{ - 1}}\operatorname{vec} \left( {{\mathbf{A}}\left( \theta  \right)} \right) \hfill \\
   \;\;\; = {\operatorname{SNR}}_R\sigma _R^2\operatorname{tr} \left( {{\mathbf{A}}\left( \theta  \right){{\mathbf{A}}^H}\left( \theta  \right){{\left( {{\mathbf{J}} + \sigma _R^2{{\mathbf{I}}_M}} \right)}^{ - 1}}} \right), \hfill \\ 
\end{gathered} 
\end{equation}
where we define radar SNR as ${\operatorname{SNR}}_R=\frac{{{{\left| \alpha  \right|}^2}L{P_R}}}{\sigma_R^2}$ \cite{1703855}.
To maintain a constant false alarm rate $P_{FA}$, $\eta$ is decided by the given $P_{FA}$ under Neyman-Pearson criterion \cite{kay1998fundamentals2}, i.e.,
\begin{equation}
    {P_{FA}} = 1 - {{\mathfrak{F}}_{{\cal X}_2^2}}\left( \eta  \right),
\end{equation}
\begin{equation}
    \eta={{\mathfrak{F}}^{-1}_{{\cal X}_2^2}}(1-{P_{FA}}),
\end{equation}
where ${{\mathfrak{F}}^{-1}_{{\cal X}_2^2}}$ is the inverse function of chi-squared Cumulative Distribution Function (CDF) with 2 DoFs. The detection probability is thus given as
\begin{equation}
{P_D} = 1-{\mathfrak{F}_{\mathcal{X}_2^2\left( \rho  \right)}}(\eta)= 1 - {\mathfrak{F}_{\mathcal{X}_2^2\left( \rho  \right)}}\left( {\mathfrak{F}_{\mathcal{X}_2^2}^{ - 1}\left( {1 - {P_{FA}}} \right)} \right),
\end{equation}
where ${\mathfrak{F}_{\mathcal{X}_2^2\left( \rho  \right)}}$ is the non-central chi-squared CDF with 2 DoFs.
\\\indent It is well-known that the accuracy of parameter estimation can be measured by the Cram\'er-Rao bound \cite{kay1998fundamentals1}, which is the lower bound for all the unbiased estimators. In our case, the parameters to be estimated are $\theta$ and $\alpha$. The Fisher Information Matrix is partitioned as
\begin{equation}
{\mathbf{{\pmb \xi} }}\left( {{\mathbf{\tilde y}}} \right) = \left[ {\begin{array}{*{20}{c}}
  {{{{{ \xi} }}_{\theta \theta }}}&{{{\mathbf{{\pmb \xi} }}_{\theta \alpha }^T}} \\ 
  {{\mathbf{{\pmb \xi} }}_{\theta \alpha }}&{{{\mathbf{{\pmb \xi} }}_{\alpha \alpha }}} 
\end{array}} \right],
\end{equation}
where $\xi_{\theta \theta }$ is a scalar, ${\pmb \xi} _{\theta \alpha }$ is a vector and  ${\pmb \xi} _{\alpha \alpha }$ is a matrix for the reason that $\theta$ is a real parameter while $\alpha$ is complex. The CRB for DoA estimation is given by
\begin{equation}
\operatorname{CRB} \left( \theta  \right) = {\left( {{{{{\xi} }}_{\theta \theta }} - {\mathbf{{\pmb \xi} }}_{\theta \alpha }^T{\mathbf{{\pmb \xi} }}_{\alpha \alpha }^{ - 1}{{\mathbf{{\pmb \xi} }}_{\theta \alpha }}} \right)^{ - 1}}.
\end{equation}
By the similar derivation as \cite{1703855}, ${\xi}_{\theta\theta}$, ${\mathbf{{\pmb \xi} }}_{\alpha \alpha }$ and ${\mathbf{{\pmb \xi} }}_{\theta \alpha }$ are given as
\begin{equation}
\begin{gathered}
{{\xi} _{\theta \theta }} = 2{\left| \alpha  \right|^2}L{P_R}\operatorname{tr} \left( {{\mathbf{\dot A}}\left( \theta  \right){{{\mathbf{\dot A}}}^H}\left( \theta  \right){{{\mathbf{\tilde J}}}^{ - 1}}} \right),\hfill \\
{{\pmb \xi} _{{\alpha \alpha }}} = 2L{P_R}\operatorname{tr} \left( {{\mathbf{A}}\left( \theta  \right){{\mathbf{A}}^H}\left( \theta  \right){{{\mathbf{\tilde J}}}^{ - 1}}} \right){{\mathbf{I}}_2},\hfill \\
{{\pmb \xi} _{\theta {\alpha }}} = 2L{P_R}\operatorname{Re} \left( {{\alpha ^*}\operatorname{tr} \left( {{\mathbf{A}}\left( \theta  \right){{{\mathbf{\dot A}}}^H}\left( \theta  \right){{{\mathbf{\tilde J}}}^{ - 1}}} \right)\left( {1;j} \right)} \right),
\end{gathered}
\end{equation}
where ${\mathbf{\dot A}}\left( \theta  \right) = \frac{{\partial {\mathbf{A}}\left( \theta  \right)}}{{\partial \theta }}$. By substituting (49) into (48), we have 
\begin{equation}
\begin{gathered}
  \operatorname{CRB} \left( \theta  \right) \hfill \\
   = \frac{1}{{2{{\operatorname{SNR} }_R}\sigma _R^2}}\times \hfill \\
   \;\;\;\;\frac{{\operatorname{tr} \left( {{\mathbf{A}}{{\mathbf{A}}^H}{{{\mathbf{\tilde J}}}^{ - 1}}} \right)}}{{\operatorname{tr} \left( {{\mathbf{\dot A}}{{{\mathbf{\dot A}}}^H}{{{\mathbf{\tilde J}}}^{ - 1}}} \right)\operatorname{tr} \left( {{\mathbf{A}}{{\mathbf{A}}^H}{{{\mathbf{\tilde J}}}^{ - 1}}} \right) - {{\left| {\operatorname{tr} \left( {{\mathbf{A}}{{{\mathbf{\dot A}}}^H}{{{\mathbf{\tilde J}}}^{ - 1}}} \right)} \right|}^2}}},
\end{gathered} 
\end{equation}
\subsection{Constructive Interference based Beamforming}
The proposed CI-based beamforming should be computed symbol by symbol, which means that the precoding vectors are functions of the time index, thus the corresponding hypothesis testing problem (35) is modified as
\begin{equation}
{{\bf{y}}^R_l}=\left\{ \begin{gathered}
  {{\cal H}_1}:\alpha \sqrt {{P_R}} {\mathbf{A}}\left( \theta  \right){\mathbf{s}}_l + {{\mathbf{G}}^T} {\tilde {\bf w}}[l]+ {\mathbf{z}}_l,\hfill \\
  \;\;\;\;\;\;\;\;\;  l = 1,2,...,L, \hfill \\
  {{\cal H}_0}:{{\mathbf{G}}^T} {\tilde {\bf w}}[l] + {\mathbf{z}}_l,\;\;l = 1,2,...,L, \hfill \\ 
\end{gathered}  \right.
\end{equation}
where ${\tilde {\bf w}}[l]={\bf w}[l]{e^{j{\phi _1}[l]}}$. While the exact analytic form of the distribution for ${\bf w}[l]$ is hard to derive, here we employ the Gaussian detector for SDR beamformer in (41). We note that for CI precoding, ${\bf w}[l]$ is not in general Gaussian, but our results show that this is indeed an affordable approximation, and, even with a Gaussian detector, CI-based beamformer achieves better performance at radar. Following the same procedure of the previous subsection, we have 
\begin{equation}
{\mathbf{J}}=\frac{1}{L}\sum\limits_{l = 1}^L{{{\mathbf{G}}^T}{\tilde {\bf w}}[l]{\tilde {\bf w}^H}[l]{{\mathbf{G}}^*}}=\frac{1}{L}\sum\limits_{l = 1}^L{{{\mathbf{G}}^T}{{\bf w}}[l]{{\bf w}^H}[l]{{\mathbf{G}}^*}} .
\end{equation}
By substituting (52) into (46) and (50) we obtain the detection probability and the $\operatorname{CRB}(\theta)$ of CI-based beamforming method.

\section{Robust Beamforming for Power Minimization with Bounded CSI Errors}
\subsection{Channel Error Model}
It is generally difficult to obtain perfect CSI in the practical scenarios. In this section, we study the beamforming design for imperfect CSI. Following the standard assumptions in the related literatures, let us first model the channel vectors as
\begin{equation}
    \begin{gathered}
  {{\mathbf{h}}_i} = {{{\mathbf{\hat h}}}_i} + {{\mathbf{e}}_{hi}}, {{\mathbf{f}}_i} = {{{\mathbf{\hat f}}}_i} + {{\mathbf{e}}_{fi}},\forall i, \hfill \\
  {{\mathbf{g}}_m} = {{{\mathbf{\hat g}}}_m} + {{\mathbf{e}}_{gm}},\forall m, \hfill \\
\end{gathered}
\end{equation}
where $ {\bf{\hat h}}_i $, $ {\bf{\hat g}}_m $ and $ {\bf{\hat f}}_i $ denote the estimated channel vectors known to the BS, $ {\bf{e}}_{hi} $, $ {\bf{e}}_{gm} $ and $ {\bf{e}}_{fi} $ denote the CSI uncertainty within the spherical sets $ {{\cal{U}}_{hi}}=\left\{ {{{\bf{e}}_{hi}}|{{\left\| {{{\bf{e}}_{hi}}} \right\|}^2} \le \delta _{hi}^2} \right\}  $, $ {{\cal{U}}_{gm}}=\left\{ {{{\bf{e}}_{gm}}|{{\left\| {{{\bf{e}}_{gm}}} \right\|}^2} \le \delta _{gm}^2} \right\}  $ and $ {{\cal{U}}_{fi}}=\left\{ {{{\bf{e}}_{fi}}|{{\left\| {{{\bf{e}}_{fi}}} \right\|}^2} \le \delta _{fi}^2} \right\}  $. This model is reasonable for scenarios that CSI is quantized at the receiver and fed back to the BS. Particularly, if the quantizer is uniform, the quantization error region can be covered by spheres of given sizes \cite{4586299}.
\\\indent It is assumed that BS has no knowledge about the error vectors except for the bounds of their norms. We therefore consider a worst-case approach to guarantee the solution is robust to all the uncertainties in above spherical sets. It should be highlighted that this is only valid when all the uncertainties lie in the constraints. For the interference minimization problem, we can not formulate a robust problem in the real sense because the uncertainty of the channel $\bf G$ lies in the objective function. However, a weighting minimization method can be applied for the case to obtain a suboptimal result. Readers are referred to \cite{liu2016robust} for details. Due to the limited space, we designate this as the objective of the future work, and focus on the robust version for power minimization in this paper.
\subsection{SDR-based Robust Beamforming}
The robust version of the SDR-based problem ${{\cal{P}}_0}$ is given by
\begin{equation}
   \begin{array}{*{20}{l}}
  {{{\cal{P}}_{10}}:\mathop {\min }\limits_{ {{{\mathbf{t}}_k}} } \;\;\sum\limits_{k = 1}^K {{{\left\| {{{\mathbf{t}}_k}} \right\|}^2}} } \\ 
\displaystyle  {s.t.\;\;\frac{{{{\left| {{\mathbf{h}}_i^T{{\mathbf{t}}_i}} \right|}^2}}}{{\sum\limits_{k = 1,k \ne i}^K {{{\left| {{\mathbf{h}}_i^T{{\mathbf{t}}_k}} \right|}^2} + {P_R}{{\left\| {{{\mathbf{f}}_i}} \right\|}^2} + \sigma _C^2} }} \ge {\Gamma _i},}
  {\text{ }} \\ 
  {\;\;\;\;\;\;\forall {{\mathbf{e}}_{hi}} \in {{\cal{U}}_{hi}},\forall {{\mathbf{e}}_{fi}} \in {{\cal{U}}_{fi}},\forall i,} \\ 
\displaystyle  {\;\;\;\;\;\;\frac{{{{\left| {{\mathbf{g}}_m^T\sum\limits_{k = 1}^K {{{\mathbf{t}}_k}}{d_k} } \right|}^2}}}{{\sigma _R^2}} \le {R_m},\forall {{\mathbf{e}}_{gm}} \in {{\cal{U}}_{gm}},\forall m.} \\ 
\end{array}
\end{equation}
The above problem is then reformulated as a worst-case approach, and can be solved by employing the well-known S-procedure \cite{boyd2004convex}. According to basic linear algebra, we have
\begin{equation}
    {{{\left\| {{{\mathbf{f}}_i}} \right\|}^2} = {{\left\| {{{{\mathbf{\hat f}}}_i} + {{\mathbf{e}}_{fi}}} \right\|}^2} \le {{\left( {\left\| {{{{\mathbf{\hat f}}}_i}} \right\| + \left\| {{{\mathbf{e}}_{fi}}} \right\|} \right)}^2} \le {{\left( {\left\| {{{{\mathbf{\hat f}}}_i}} \right\| + {\delta _{fi}}} \right)}^2}.}
\end{equation}
Similarly, for the interference power we have
\begin{equation}
\begin{gathered}
  {\left| {{\mathbf{g}}_m^T\sum\limits_{k = 1}^K {{{\mathbf{t}}_k}} {d_k}} \right|^2} = \sum\limits_{k = 1}^K {\operatorname{tr} \left( {\left( {{\mathbf{\hat g}}_m^* + {\mathbf{e}}_{gm}^*} \right)\left( {{\mathbf{\hat g}}_m^T + {\mathbf{e}}_{gm}^T} \right){{\mathbf{t}}_k}{\mathbf{t}}_k^H} \right)}  \hfill \\
   = \sum\limits_{k = 1}^K {\operatorname{tr} \left( {\left( {{\mathbf{\hat g}}_m^*{\mathbf{\hat g}}_m^T + {\mathbf{\hat g}}_m^*{\mathbf{e}}_{gm}^T + {\mathbf{e}}_m^*{\mathbf{\hat g}}_m^T + {\mathbf{e}}_{gm}^*{\mathbf{e}}_{gm}^T} \right){{\mathbf{t}}_k}{\mathbf{t}}_k^H} \right)}.  \hfill \\ 
\end{gathered} 
\end{equation}
By using the Cauchy-Schwarz inequality and rearranging the formula, it follows that
\begin{equation}
\begin{gathered}
  {\left| {{\mathbf{g}}_m^T\sum\limits_{k = 1}^K {{{\mathbf{t}}_k}}{d_k} } \right|^2} \le 
   \sum\limits_{k = 1}^K {\operatorname{tr} \left( {{\mathbf{\hat g}}_m^*{\mathbf{\hat g}}_m^T{{\mathbf{t}}_k}{\mathbf{t}}_k^H} \right)} {\kern 1pt}  \hfill \\
   + \left( {2\left\| {{{{\mathbf{\hat g}}}_m}} \right\|\left\| {{{\mathbf{e}}_{gm}}} \right\| + {{\left\| {{{\mathbf{e}}_{gm}}} \right\|}^2}} \right)\sum\limits_{k = 1}^K {\operatorname{tr} \left( {{{\mathbf{t}}_k}{\mathbf{t}}_k^H} \right)}  \hfill \\
   \le \sum\limits_{k = 1}^K {\operatorname{tr} \left( {{\mathbf{\hat g}}_m^*{\mathbf{\hat g}}_m^T{{\mathbf{t}}_k}{\mathbf{t}}_k^H} \right)} {\kern 1pt}  + 
  \left( {2{\delta _{gm}}\left\| {{{{\mathbf{\hat g}}}_m}} \right\| + \delta _{gm}^2} \right)\sum\limits_{k = 1}^K {\operatorname{tr} \left( {{{\mathbf{t}}_k}{\mathbf{t}}_k^H} \right)}.  \hfill \\ 
\end{gathered} 
\end{equation}
Based on the work \cite{liu2016robust}, we directly give the worst-case formulation of ${\cal{P}}_{10}$ by 
\begin{equation}
\begin{gathered}
  {{\cal P}_{11}}:\mathop {\min }\limits_{{{\mathbf{T}}_i},{s_i}} \sum\limits_{i = 1}^K {\operatorname{tr} \left( {{{\mathbf{T}}_i}} \right)}  \hfill \\
  s.t.\;\left[ {\begin{array}{*{20}{c}}
  {{\mathbf{\hat h}}_i^T{{\mathbf{Q}}_i}{{{\mathbf{\hat h}}}_i^*} - {\Gamma _i}{\beta _i} - {s_i}\delta _{hi}^2}&{{\mathbf{\hat h}}_i^T{{\mathbf{Q}}_i}} \\ 
  {{{\mathbf{Q}}_i}{{{\mathbf{\hat h}}}_i^*}}&{{{\mathbf{Q}}_i} + s _i{\mathbf{I}}} 
\end{array}} \right] \succeq 0, \hfill \\
  \;\;\;\;\;\;\;{{\mathbf{T}}_i} \succeq 0,{{\mathbf{T}}_i} = {\mathbf{T}}_i^*,{\text{rank}}\left( {{{\mathbf{T}}_i}} \right) = 1,{s_i} \ge 0,\forall i, \hfill \\
  \;\;\;\;\;\;\;\sum\limits_{i = 1}^K {\left( \begin{gathered}
  \operatorname{tr} \left( {{{{\mathbf{\hat g}}}_m}^*{\mathbf{\hat g}}_m^T{{\mathbf{T}}_i}} \right) 
   + {\zeta_{gm}}\operatorname{tr} \left( {{{\mathbf{T}}_i}} \right) \hfill \\ 
\end{gathered}  \right)}  \le {R_m}\sigma _R^2,\forall m, \hfill \\ 
\end{gathered} 
\end{equation}
where $ {{\bf{T}}_k}={{\bf{t}}_k}{\bf{t}}_k^H $, $ {{\mathbf{Q}}_i} = {{\mathbf{T}}_i} - {\Gamma _i}\sum\limits_{n = 1,n \ne i}^K {{{\mathbf{T}}_n}}  $, $ {\zeta_{gm}}={2{\delta _2}\left\| {{{{\mathbf{\hat g}}}_m}} \right\| + \delta _{gm}^2}  $ and $ {{\beta _i}={P_R}{{\left( {\left\| {{{{\mathbf{\hat f}}}_i}} \right\| + {\delta _{fi}}} \right)}^2} + \sigma _C^2} $. By dropping the rank constraints on ${\bf T}_i$, the above problem becomes a standard SDP and can be solved by SDR method, after which the beamforming vectors can be obtained by rank-1 approximation or Gaussian randomization \cite{5447068}.
\subsection{Constructive Interference based Robust Beamforming}
Let us first formulate the robust version of the virtual multicast problem ${{\cal P}_{3}}$ as
\begin{equation}
\begin{array}{*{20}{l}}
  {{\cal{P}}_{12}}:\mathop {\min }\limits_{\mathbf{w}} \;\;{\kern 1pt} {\kern 1pt} {\left\| {\mathbf{w}} \right\|^2} \hfill \\
 \displaystyle s.t.{\kern 1pt}\left| {\operatorname{Im} \left( {{\mathbf{{\tilde h}}}_i^T{\mathbf{w}}} \right)} \right| \le \left( {\operatorname{Re} \left( {{\mathbf{{\tilde h}}}_i^T{\mathbf{w}}} \right) - \sqrt {{{\tilde \Gamma }_i}} } \right)\tan {\psi }, \hfill \\
 {\;\;\;\;\;\;\forall {{\mathbf{e}}_{hi}} \in {{\cal{U}}_{hi}},\forall {{\mathbf{e}}_{fi}} \in {{\cal{U}}_{fi}},\forall i,} \\ 
  \;\;\;\;\;{\kern 1pt}\left| {{\mathbf{\tilde g}}_m^T{\mathbf{w}}} \right| \le \sqrt {{R_m}\sigma _R^2}, \forall {{\mathbf{e}}_{gm}} \in {{\cal{U}}_{gm}},\forall m.\\ 
\end{array}
\end{equation}
Apparently the robust case for the channel vector ${\bf f}_i$ is the same as (55). Consider the worst case of the INR constraints, which is 
\begin{equation}
\max\;\;\left| {{\mathbf{\tilde g}}_m^T{\mathbf{w}}} \right| \le \sqrt {{R_m}\sigma _R^2}, \forall {{\mathbf{e}}_{gm}} \in {{\cal{U}}_{gm}},\forall m.
\end{equation}
Since ${{\mathbf{\tilde g}}_m} \triangleq {{\mathbf{g}}_m}{e^{j{\phi _1}}}$, it is easy to see ${\left\| {{{{\mathbf{\tilde g}}}_m}{\mathbf{w}}} \right\|^2} = {\left\| {{{\mathbf{g}}_m}{\mathbf{w}}} \right\|^2}$. For the convenience of further analysis, we drop the subscript, and denote the interference channel vector by its real and imaginary parts, which is given by
\begin{equation}
{\mathbf{g}} = {{{{\mathbf{\hat g}}}_R} + j{{{\mathbf{\hat g}}}_I} + {{\mathbf{e}}_{gR}} + j{{\mathbf{e}}_{gI}}}.
\end{equation}
Let ${\mathbf{\bar g}} = \left[ {{{{\mathbf{\hat g}}}_{R}};{{{\mathbf{\hat g}}}_{I}}} \right], {{{\mathbf{\bar e}}}_g} = \left[ {{{\mathbf{e}}_{gR}};{{\mathbf{e}}_{gI}}} \right]$, the interference from radar can be written as
\begin{equation}
\begin{gathered}
  {\left| {{\mathbf{\tilde g}}^T{\mathbf{w}}} \right|^2} = {\left\| {\left[ {\begin{array}{*{20}{c}}
  {{\mathbf{\hat g}}_R^T + {\mathbf{e}}_{gR}^T}&{{\mathbf{\hat g}}_I^T + {\mathbf{e}}_{gI}^T} \\ 
  {{\mathbf{\hat g}}_I^T + {\mathbf{e}}_{gI}^T}&{ - {\mathbf{\hat g}}_R^T - {\mathbf{e}}_{gR}^T} 
\end{array}} \right]\left[ {\begin{array}{*{20}{c}}
  {{{\mathbf{w}}_R}} \\ 
  { - {{\mathbf{w}}_I}} 
\end{array}} \right]} \right\|^2} \hfill \\
   = {\left\| \begin{gathered}
  {{{\mathbf{\bar g}}}^T}{{\mathbf{w}}_2} + {\mathbf{\bar e}}_g^T{{\mathbf{w}}_2} \hfill \\
  {{{\mathbf{\bar g}}}^T}{{\mathbf{w}}_1} + {\mathbf{\bar e}}_g^T{{\mathbf{w}}_1} \hfill \\ 
\end{gathered}  \right\|^2}, \hfill \\ 
\end{gathered}
\end{equation}
According to the Cauchy-Schwarz inequality, (62) can be further expanded as
\begin{equation}
\begin{gathered}
  {\left\| \begin{gathered}
  {{{\mathbf{\bar g}}}^T}{{\mathbf{w}}_2} + {\mathbf{\bar e}}_g^T{{\mathbf{w}}_2} \hfill \\
  {{{\mathbf{\bar g}}}^T}{{\mathbf{w}}_1} + {\mathbf{\bar e}}_g^T{{\mathbf{w}}_1} \hfill \\ 
\end{gathered}  \right\|^2} \hfill \\
   \le {\left| {{{{\mathbf{\bar g}}}^T}{{\mathbf{w}}_2}} \right|^2} + {\left| {{{{\mathbf{\bar g}}}^T}{{\mathbf{w}}_1}} \right|^2} + 2\delta _g^2{\left\| {{{\mathbf{w}}_2}} \right\|^2} \hfill \\
   \quad + 2{\delta _g}\left( {\left\| {{{{\mathbf{\bar g}}}^T}{{\mathbf{w}}_2}{\mathbf{w}}_2^T} \right\| + \left\| {{{{\mathbf{\bar g}}}^T}{{\mathbf{w}}_1}{\mathbf{w}}_1^T} \right\|} \right) \hfill \\
   \le {\left| {{{{\mathbf{\bar g}}}^T}{{\mathbf{w}}_2}} \right|^2} + {\left| {{{{\mathbf{\bar g}}}^T}{{\mathbf{w}}_1}} \right|^2} + \left( {2\delta _g^2 + 4{\delta _g}\left\| {{\mathbf{\bar g}}} \right\|} \right){\left\| {{{\mathbf{w}}_2}} \right\|^2}, \hfill \\ 
\end{gathered}
\end{equation}
and the robust constraint for INR is given by
\begin{equation}
{\left| {{{{\mathbf{\bar g}}}^T}{{\mathbf{w}}_2}} \right|^2} + {\left| {{{{\mathbf{\bar g}}}^T}{{\mathbf{w}}_1}} \right|^2} + \left( {2\delta _g^2 + 4{\delta _g}\left\| {{\mathbf{\bar g}}} \right\|} \right){\left\| {{{\mathbf{w}}_2}} \right\|^2}\le R\sigma _R^2. 
\end{equation}
For the SINR constraint, note that the corresponding worst case is equivalent to 
\begin{equation}
\begin{gathered}
  \max\;\;\left| {\operatorname{Im} \left( {{\mathbf{{\tilde h}}}_i^T{\mathbf{w}}} \right)} \right| - \operatorname{Re} \left( {{\mathbf{{\tilde h}}}_i^T{\mathbf{w}}} \right)\tan \psi  + \sqrt {{{\tilde \Gamma }_i}} \tan \psi  \le 0, \hfill \\
 \;\;\;\;\;\;\;\;\;\;\forall {{\mathbf{e}}_{hi}} \in {{\cal{U}}_{hi}},\forall {{\mathbf{e}}_{fi}} \in {{\cal{U}}_{fi}},\forall i. \hfill \\ 
\end{gathered}
\end{equation}
Let ${{\mathbf{\hat {{\tilde h}}}}_i} = {{\mathbf{\hat h}}_i}{e^{j\left( {{\phi _1} - {\phi _i}} \right)}},{{\mathbf{\tilde e}}_{hi}} = {{\mathbf{e}}_{hi}}{e^{j\left( {{\phi _1} - {\phi _i}} \right)}}$, we have ${{{\mathbf{{\tilde h}}}}_i} = {{{\mathbf{\hat {{\tilde h}}}}}_i} + {{{\mathbf{\tilde e}}}_{hi}}$. Similarly, we drop the subscript and denote the channel vector by its real and imaginary parts, which is 
\begin{equation}
{\mathbf{{\tilde h}}} = {{{\mathbf{\hat {{\tilde h}}}}}_R} + j{{{\mathbf{\hat {{\tilde h}}}}}_I} + {{{\mathbf{\tilde e}}}_{hR}} + j{{{\mathbf{\tilde e}}}_{hI}}.
\end{equation}
It follows that
\begin{equation}
\begin{gathered}
  \operatorname{Im} \left( {{\mathbf{{\tilde h}w}}} \right) = \operatorname{Im} \left( {\left( {{{{\mathbf{\hat {\tilde h}}}}_R} + j{{{\mathbf{\hat {\tilde h}}}}_I} + {{{\mathbf{\tilde e}}}_{hR}} + j{{{\mathbf{\tilde e}}}_{hI}}} \right)\left( {{{\mathbf{w}}_R} + j{{\mathbf{w}}_I}} \right)} \right) \hfill \\
   = \left[ {{{{\mathbf{\hat {\tilde h}}}}_R},{{{\mathbf{\hat {\tilde h}}}}_I}} \right]\left[ \begin{gathered}
  {{\mathbf{w}}_I} \hfill \\
  {{\mathbf{w}}_R} \hfill \\ 
\end{gathered}  \right] + \left[ {{{{\mathbf{\tilde e}}}_{hR}},{{{\mathbf{\tilde e}}}_{hI}}} \right]\left[ \begin{gathered}
  {{\mathbf{w}}_I} \hfill \\
  {{\mathbf{w}}_R} \hfill \\ 
\end{gathered}  \right] \hfill \\
   \triangleq {{{\mathbf{\hat {{\bar h}}}}}^T}{{\mathbf{w}}_1} + {\mathbf{\bar e}}_h^T{{\mathbf{w}}_1}, \hfill \\ 
\end{gathered} 
\end{equation}

\begin{equation}
\begin{gathered}
  \operatorname{Re} \left( {{\mathbf{{\tilde h}w}}} \right) = \operatorname{Re} \left( {\left( {{{{\mathbf{\hat {\tilde h}}}}_R} + j{{{\mathbf{\hat {\tilde h}}}}_I} + {{{\mathbf{\tilde e}}}_{hR}} + j{{{\mathbf{\tilde e}}}_{hI}}} \right)\left( {{{\mathbf{w}}_R} + j{{\mathbf{w}}_I}} \right)} \right) \hfill \\
   = \left[ {{{{\mathbf{\hat {\tilde h}}}}_R},{{{\mathbf{\hat {\tilde h}}}}_I}} \right]\left[ \begin{gathered}
  {{\mathbf{w}}_R} \hfill \\
  {-{\mathbf{w}}_I} \hfill \\ 
\end{gathered}  \right] + \left[ {{{{\mathbf{\tilde e}}}_{hR}},{{{\mathbf{\tilde e}}}_{hI}}} \right]\left[ \begin{gathered}
  {{\mathbf{w}}_R} \hfill \\
  {-{\mathbf{w}}_I} \hfill \\ 
\end{gathered}  \right] \hfill \\
   \triangleq {{{\mathbf{\hat {{\bar h}}}}}^T}{{\mathbf{w}}_2} + {\mathbf{\bar e}}_h^T{{\mathbf{w}}_2}. \hfill \\ 
\end{gathered} 
\end{equation}
By noting that ${\left\| {{{{\mathbf{\bar e}}}_h}} \right\|^2} \le \delta _h^2$, (65) is equivalent to 
\begin{equation}
\begin{gathered}
  \max \left| {{{{\mathbf{\hat {{\bar h}}}}}^T}{{\mathbf{w}}_1} + {\mathbf{\bar e}}_h^T{{\mathbf{w}}_1}} \right| - \left( {{{{\mathbf{\hat {{\bar h}}}}}^T}{{\mathbf{w}}_2} + {\mathbf{\bar e}}_h^T{{\mathbf{w}}_2}} \right)\tan \psi  \hfill \\
  \;\;\;\;\;\;\; + \sqrt {\tilde \Gamma } \tan \psi  \le 0, \forall {\left\| {{{{\mathbf{\bar e}}}_h}} \right\|^2} \le \delta _h^2,\forall {\left\| {{{\mathbf{e}}_f}} \right\|^2} \le \delta _f^2,
\end{gathered}
\end{equation}
and can be decomposed into the following two constraints:
\begin{equation}
\begin{gathered}
  \max \;{{{{\mathbf{\hat {{\bar h}}}}}^T}{{\mathbf{w}}_1} + {\mathbf{\bar e}}_h^T{{\mathbf{w}}_1}}  - \left( {{{{\mathbf{\hat {{\bar h}}}}}^T}{{\mathbf{w}}_2} + {\mathbf{\bar e}}_h^T{{\mathbf{w}}_2}} \right)\tan \psi  \hfill \\
  \;\;\;\;\;\;\; + \sqrt {\tilde \Gamma } \tan \psi  \le 0, \forall {\left\| {{{{\mathbf{\bar e}}}_h}} \right\|^2} \le \delta _h^2,\forall {\left\| {{{\mathbf{e}}_f}} \right\|^2} \le \delta _f^2,
\end{gathered}
\end{equation}
\begin{equation}
\begin{gathered}
  \max -{{{{\mathbf{\hat {{\bar h}}}}}^T}{{\mathbf{w}}_1} -{\mathbf{\bar e}}_h^T{{\mathbf{w}}_1}}  - \left( {{{{\mathbf{\hat {{\bar h}}}}}^T}{{\mathbf{w}}_2} + {\mathbf{\bar e}}_h^T{{\mathbf{w}}_2}} \right)\tan \psi  \hfill \\
  \;\;\;\;\;\;\; + \sqrt {\tilde \Gamma } \tan \psi  \le 0, \forall {\left\| {{{{\mathbf{\bar e}}}_h}} \right\|^2} \le \delta _h^2,\forall {\left\| {{{\mathbf{e}}_f}} \right\|^2} \le \delta _f^2. 
\end{gathered}
\end{equation}
Based on above, the worst-case constraints for (70) and (71) are given by
\begin{equation}
\begin{gathered}
  {{{\mathbf{\hat {\bar h}}}}^T}{{\mathbf{w}}_1} - {{{\mathbf{\hat {\bar h}}}}^T}{{\mathbf{w}}_2}\tan \psi  + {\delta _h}\left( {{{\mathbf{w}}_1} - {{\mathbf{w}}_2}\tan \psi } \right) \hfill \\
  \;\; + \sqrt {\Gamma \left( {\sigma _C^2 + {P_R}\left( {{{\left\| {\mathbf{\hat f}} \right\|}} + \delta _f} \right)^2} \right)} \tan \psi  \le 0,
\end{gathered}
\end{equation}
\begin{equation}
\begin{gathered}
   - {{{\mathbf{\hat {\bar h}}}}^T}{{\mathbf{w}}_1} - {{{\mathbf{\hat {\bar h}}}}^T}{{\mathbf{w}}_2}\tan \psi  + {\delta _h}\left( {{{\mathbf{w}}_1} + {{\mathbf{w}}_2}\tan \psi } \right) \hfill \\
  \;\;\;\;\;\;+ \sqrt {\Gamma \left( {\sigma _C^2 + {P_R}\left( {{{\left\| {\mathbf{\hat f}} \right\|}} + \delta _f} \right)^2} \right)} \tan \psi  \le 0. \hfill \\ 
\end{gathered}
\end{equation}
\\\indent The final robust optimization problem is given by
\begin{equation}
\begin{array}{*{20}{l}}
  {{\cal{P}}_{13}}:\mathop {\min }\limits_{{\mathbf{w}}_1,{\mathbf{w}}_2} \;\;{\kern 1pt} {\kern 1pt} {\left\| {{\mathbf{w}}_1} \right\|^2} \hfill \\
  s.t.\;\;\;\; \text{Constraints (64), (72) and (73)}, \forall i, \forall m, \hfill \\
  \;\;\;\;\;\;\;\;\;{{\mathbf{w}}_1} = {\mathbf{\Pi }}{{\mathbf{w}}_2}.
\end{array}
\end{equation}

\section{Numerical Results}
In this section, numerical results based on Monte Carlo simulations are shown to validate the effectiveness of the proposed beamforming method. Without loss of generality, all the channel matrices follow the standard complex Gaussian distribution, and are independent and identically distributed (i.i.d). For simplicity, the INR thresholds for different radar antennas, the SINR level for different downlink users and the error bounds for different channels are set to be equal, respectively, i.e., $ R_m = R, \Gamma_i = \Gamma, \delta_{hi} = \delta_{fi} = \delta_{gm} = \delta, \forall i, \forall m $. While it is plausible that the benefits of the proposed scheme extend to various scenarios, here we assume $\alpha=P_R=1$, $ N=8 $, $ K=M=4 $ unless otherwise specified, and explore the results for QPSK and 8PSK modulations. The power of all the noise vectors are set to be 1mW, i.e., ${\sigma_R^2} = {\sigma_C^2} = 0\text{dBm}$. We denote the conventional SDR beamformer as `SDR' in the figures, and the proposed beamformer based on constructive interference as `CI'.
\subsection{Average Transmit Power}
In Fig. 3, we compare the minimized power for the two beamforming methods under a given INR level of 0dB with the increasing $\Gamma$. Unsurprisingly, the power needed for transmission increases with growing $\Gamma$ for both methods. However, it can be easily seen that the proposed method obtains a lower transmit power for given INR and SINR requirements than the conventional SDR-based method thanks to the exploitation of the constructive interference. Particularly if QPSK modulation is used, the required power for CI-based scheme is less than half of the power needed for SDR-based beamforming. Similar results have been provided in Fig. 4, where the transmit power of different methods with increased $R$ has been given with required SINR fixed at 20dB and 24dB respectively. It is worth noting that there exists a trade-off between the power needed for BS and the INR level received by radar as has been discussed in the previous section. For both SINR levels, the proposed method performs far better than the conventional one especially in all practical INR levels.
\begin{figure}
    \centering
    \includegraphics[width=3.0in]{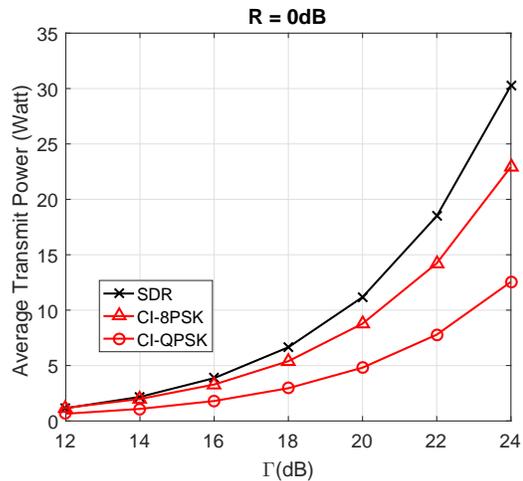}
    \caption{Average transmit power vs. required SINR, with $R = 0\text{dB}$.}
    \label{fig:3}
\end{figure}
\begin{figure}
    \centering
    \includegraphics[width=3.0in]{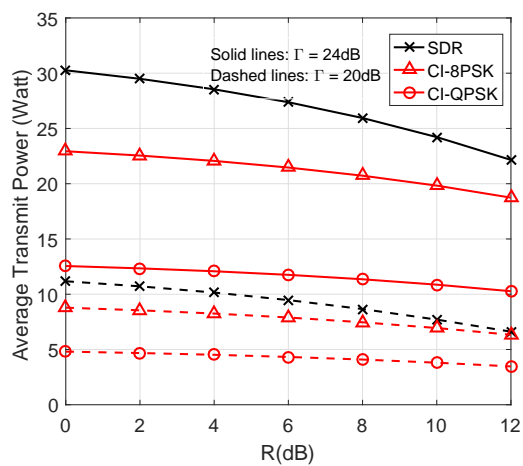}
    \caption{Trade-off between BS transmit power and INR level, with $\Gamma = 20\text{dB}$ and 24dB respectively.}
    \label{fig:4}
\end{figure}
\begin{figure}
    \centering
    \includegraphics[width=3.0in]{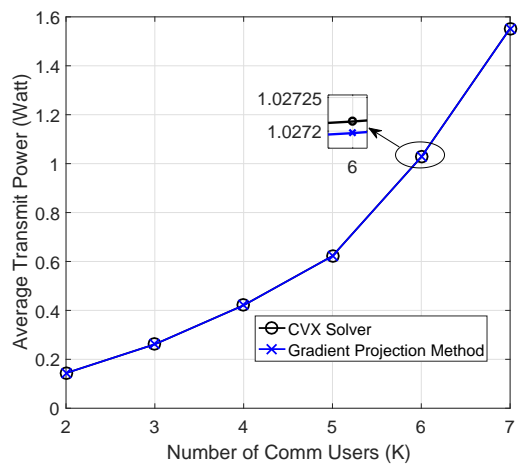}
    \caption{Result comparison of Algorithm 1 and CVX-Solver for ${\cal P}_3$, $N = 12, M = 4$, $\Gamma = 20\text{dB}, R=5\text{dB}$, QPSK.}
    \label{fig:5}
\end{figure}
\subsection{Complexity}
 In order to verify the effectiveness of the proposed efficient algorithm for ${\cal P}_3$, we compare the results obtained by the built-in \emph{SeDuMi} solver in CVX \cite{grant2008cvx} and Algorithm 1 with increasing downlink users $K$ in Fig. 5, where $N=12, M=4, \Gamma = 20\text{dB}, R=5\text{dB}$. As we can see that the two curves match very well and the difference is less than 0.05mW when $M = 6$. Since it is difficult to analytically derive the complexity of the optimization based beamforming as well as the proposed iteration algorithm, the complexity for ${\cal P}_0$, ${\cal P}_3$ and Algorithm 1 has been compared in terms of average execution time for a growing number of downlink users in Fig. 6, with the same configuration of Fig. 5. Note that it takes less time to solve ${\cal P}_3$ than ${\cal P}_0$ by the CVX solver. This is because the latter needs a rank-1 approximation or Gaussian randomization to obtain the optimal beamforming vectors, which involves extra amount of computations \cite{5447068}. Nevertheless, the proposed CI-based approach is a symbol-level beamformer, which means that the beamforming vectors should be calculated symbol by symbol while the SDR-based beamforming needs only one-time calculation during a communication frame in slow fading channels. Fortunately, the proposed gradient projection algorithm is far more efficient than the CVX solver and saves nearly 90\% of time with respect to the SDR beamformer. In a typical LTE system with 14 symbols in one frame, the total execution time for the gradient projection algorithm will be 140\% ($(1-90\%)\times 14=140\%$) of the SDR-based beamforming, but the gain of the saved transmit power is more than 200\% as has been shown in Fig. 3 and Fig. 4, which is cost-effective in energy-limited systems.

\begin{figure}
    \centering
    \includegraphics[width=3.0in]{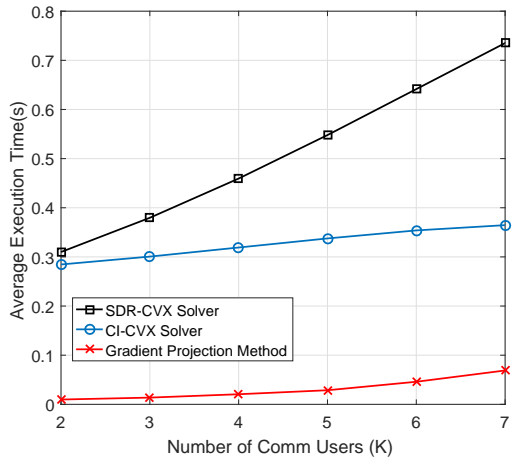}
    \caption{Average execution time for optimization ${\cal P}_0$, ${\cal P}_3$ and Algorithm 1, $N = 12, M = 4$, $\Gamma = 20\text{dB}, R=5\text{dB}$, QPSK.}
    \label{fig:6}
\end{figure} 
\begin{figure}
    \centering
    \includegraphics[width=3.0in]{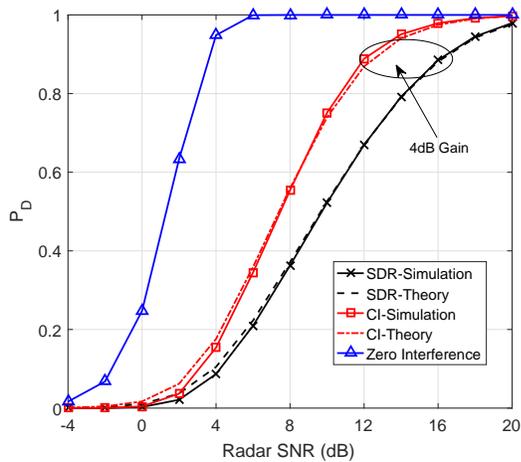}
    \caption{Detection probability vs. radar SNR for different cases, $P = 24\text{dBm}$, $\Gamma = 18\text{dB}$, $\eta=13.5\text{dBm}$, QPSK.}
    \label{fig:7}
\end{figure}
\begin{figure}
    \centering
    \includegraphics[width=3.0in]{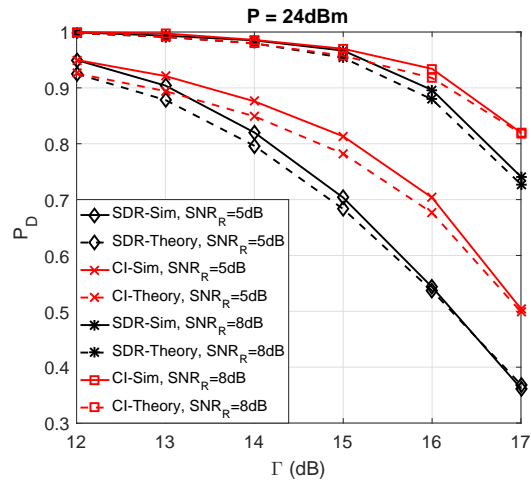}
    \caption{Detection Probability vs. SINR threshold for different cases, $P = 24\text{dBm}$, QPSK.}
    \label{fig:8}
\end{figure}
\begin{figure}
    \centering
    \includegraphics[width=3.0in]{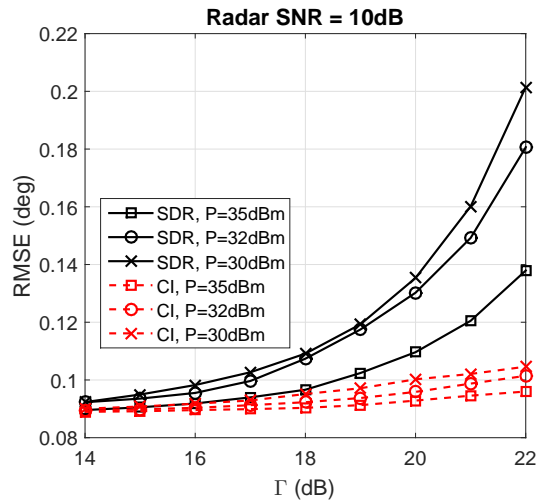}
    \caption{RMSE vs. SINR threshold for different power budget, ${\text{SNR}}_R=10\text{dB}$, QPSK.}
    \label{fig:9}
\end{figure}
\subsection{Radar Performance}
Fig. 7-9 demonstrate a series of results for the impact of the proposed scheme on different radar metrics by solving the interference minimization problem ${\cal P}_1$ and ${\cal P}_4$. Here we assume that radar is equipped with an Uniform Linear Array (ULA) with half-wavelength spacing, and \emph{m}-sequences are used as the radar waveform with a length of 40 digits, i.e., $L=40$. The target is set to be located at the direction of $\theta=\pi/5$.  
\\\indent In Fig. 7, the average detection probability with increased radar SNR for the two methods are given, where the solid line with triangle markers denotes the case without interference from the BS. Among the rest lines, the solid curves and dashed ones denote the simulated and asymptotic detection performance respectively. The parameters are given as $\eta=13.5\text{dBm}$.  $\Gamma = 18\text{dB}$, and $P=24\text{dBm}$. As shown in the figure, the simulated results match well with the asymptotic ones for both SDR and CI methods. Once again, we see that the proposed method outperforms the SDR-based method significantly. For instance, the extra gain needed for the SDR method is 4dB compared with the proposed method for a desired $P_D = 0.9$.
\\\indent Fig. 8 shows another important trade-off between radar and communication, where the detection probability at the radar with increased SINR threshold of the downlink users are provided for the two methods with the same parameter configuration as Fig. 7. It can be seen that a higher SINR requirement at users leads to a lower $P_D$ for radar, and the proposed method obtains better trade-off curves for both simulated and asymptotic results thanks to the utilization of MUI. The results in Figs. 7 and 8 justify the use of the Gaussian radar detector of (41) for the CI beamformer, which still gives significant performance gains w.r.t the SDR beamformer.
\\\indent In Fig. 9, the lower bound of radar DoA estimation is given in terms of the root-mean-square-error (RMSE) with increased SINR threshold and different BS power budget, where $\text{RMSE}(\theta)=\sqrt{\text{CRB}(\theta)}$. As expected, the loose of the communication constraints brings benefits to radar target estimation. It can be also observed that the proposed approach is not only robust to the increasing SINR requirement, but also performs far better than the SDR method.

\subsection{Robust Designs}
In Fig. 10, the BS transmit power with increasing CSI error bound $\delta$ is shown with $\Gamma = 25\text{dB}, R=30\text{dB}$, where different cases with perfect and imperfect CSI are simulated for both SDR and CI-based beamforming. The legend denotes the channel which suffers from CSI errors for each case, while the rest are assumed perfectly known. Thanks to its relaxed nature, the CI-based beamforming has a higher degree of tolerance for the CSI errors than SDR-based ones. The same trend is also shown in Fig. 11, where we apply a fixed channel error bound $\delta^2=2\times 10^{-4}$ and $R=25\text{dB}$ for all the robust cases to see the variation of the transmit power with an increased SINR level. Since the interference channel between radar and users should first be estimated by the users and then fed back to the BS, the knowledge about $\bf F$ is more likely to be known inaccurately by the BS compared with other two channels. Fortunately, we observe that in both Fig. 10 and Fig. 11, the imperfect channel $\bf F$ requires less transmit power to meet the same SINR level than $\bf H$ and $\bf G$ with CSI errors of the same bound. Hence, the accuracy for the estimation of $\bf F$ can be relatively lower than the other channels.

\begin{figure}
    \centering
    \includegraphics[width=2.8in]{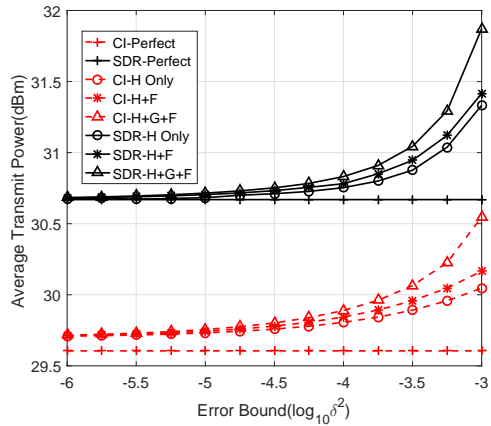}
    \caption{Average transmit power vs. error bound for different robust cases, $\Gamma = 25\text{dB}, R = 30\text{dB}$, QPSK.}
    \label{fig:10}
\end{figure}
\begin{figure}
    \centering
    \includegraphics[width=2.8in]{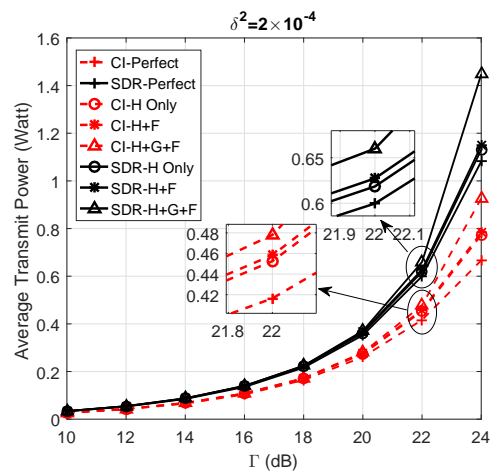}
    \caption{Average transmit power vs. SINR for different robust cases, $\delta^2=2 \times 10^{-4}, R = 25\text{dB}$, QPSK.}
    \label{fig:11}
\end{figure}

\section{Conclusion}
This paper proposes a novel optimization-based beamforming approach for MIMO radar and downlink MU-MISO communication coexistence, where multi-user interference is utilized to enhance the performance of communication system and relax the constraints in the optimization problems. Numerical results show that the proposed scheme outperforms the conventional SDR-based beamformers in terms of both power and interference minimization. An efficient gradient projection method is further given to solve the proposed power minimization problem, and is compared with SDR-based solver in the sense of average execution time. While the proposed technique is applied at symbol level, the computation complexity is still comparable with the SDR approach in typical LTE systems. Moreover, the detection probability and the Cram\'er-Rao bound for MIMO radar in the presence of the interference from BS are analytically derived, and the trade-off between the performance of radar and communication is revealed. Finally, a robust beamformer for power minimization is designed for imperfect CSI cases based on interference exploitation, and obtains significant performance gains compared with conventional schemes.

%



\section*{Acknowledgment}

This research was supported by the Engineering and Physical Sciences Research Council (EPSRC) project EP/M014150/1 and the China Scholarship Council (CSC) under Grant No. 201606030054.

\ifCLASSOPTIONcaptionsoff
  \newpage
\fi



\bibliographystyle{IEEEtran}
\bibliography{IEEEabrv,TSP_CI}
\end{document}